\documentclass[a4paper,12pt]{article}
\usepackage{amsmath}
\usepackage{graphicx}
\usepackage{amsfonts}

\DeclareFontFamily{U}{rsf}{}
\DeclareFontShape{U}{rsf}{m}{n}{
  <5> <6> rsfs5 <7> <8> <9> rsfs7 <10-> rsfs10}{}
\DeclareMathAlphabet\Scr{U}{rsf}{m}{n}

\mathchardef\varGamma="0100
\mathchardef\varDelta="0101
\mathchardef\varTheta="0102
\mathchardef\varLambda="0103
\mathchardef\varXi="0104
\mathchardef\varPi="0105
\mathchardef\varSigma="0106
\mathchardef\varUpsilon="0107
\mathchardef\varPhi="0108
\mathchardef\varPsi="0109
\mathchardef\varOmega="010A

\title{Non-perturbative Vacuum Destabilization and D-brane Dynamics}
\begin{document}

\thispagestyle{empty}

\vspace{-3.5cm}

\begin{flushright}
{\small
CPHT-RR028.0310 \\
CERN-PH-TH-2010-073 \\
LPT-ORSAY 10-24 \\

}
\end{flushright}

\begin{center}
{\bf\Large Non-perturbative Vacuum Destabilization

and D-brane Dynamics\\[18pt]}

\vspace{0.6cm}

{\bf
P. G. C\'amara$^{*}$,
C. Condeescu$^{+,}\footnote{On leave of absence from Institute of
Mathematics `Simion Stoilow' of the Romanian
Academy,\\
P.O. Box 1-764, Bucharest, RO-70700, Romania.}$,
E. Dudas$^{+,\dag}$
and M. Lennek$^{++}$  \vspace{0.1cm}
}

{\it $^*$ Theory Group CERN, 1211 Geneva 23, Switzerland\\[2pt]}
{\it $^{+}$ Centre de Physique Th\'eorique,
Ecole Polytechnique and CNRS,\\
F-91128 Palaiseau, France.\\[2pt]
$^{\dag}$ LPT,Bat. 210, Univ. de Paris-Sud, F-91405 Orsay, France
\\[2pt]
{\it $^{++}$ Dept. of Physics and Astronomy, McMaster Univ,Hamilton ON, Canada,L8S4M1 \\[2pt]}
}

{\bf Abstract}
\end{center}
We analyze the process of string vacuum destabilization due to
instanton induced superpotential couplings which depend linearly on
charged fields. These non-perturbative instabilities
result in potentials for the D-brane moduli and lead to
processes of D-brane recombination, motion and partial moduli
stabilization at the non-perturbative vacuum. By using
techniques of D-brane instanton calculus, we explicitly compute this
scalar potential in
toroidal orbifold compactifications with magnetized D-branes by
summing over the possible discrete instanton configurations. We
illustrate explicitly the resulting dynamics in globally consistent
models.
These instabilities can have phenomenological
applications to breaking hidden
sector gauge groups, open string moduli stabilization and
supersymmetry breaking. Our results suggest that breaking
supersymmetry by Polonyi-like models in string theory is
more difficult than expected.

\clearpage

\tableofcontents


\section{Introduction}
Non-perturbative effects are crucial for understanding various
problems in Quantum Field Theory (QFT) and String Theory. Instantons,
in particular, provide an elegant explanation for the spontaneous
breaking of classical symmetries in QFT through
dynamical generation of condensates. Particularly well-understood
examples are gaugino condensation induced by
fractional gauge instantons in supersymmetric theories, leading to
spontaneous breaking of R-symmetry, or the chiral
condensate induced by t'Hooft instantons in QCD, leading to spontaneous breaking of chiral symmetry. Hence, instanton effects are useful to illuminate many problems related to the vacuum structure of QCD, supersymmetry breaking or technicolor.

In String Theory the landscape of possible instanton effects becomes more intricate. There are two qualitatively different effects. The first is just the string theory description of QFT instantons: Euclidean D$(p-4)$-branes (also dubbed E$(p-4)$-branes) wrapping the same cycle in the compact space as a space-time filling stack of D$p$-branes realize QFT instanton effects for the gauge theory in the worldvolume of the D$p$-brane. Qualitatively different type of effects are generated by E$q$-branes wrapping a different cycle in the internal space and intersecting the D$p$-brane under consideration, provided that the brane and the instanton are mutually BPS. Such effects, called {\it stringy instanton effects}, arise even for Abelian gauge groups and are therefore not easily understandable in the language of gauge theory instantons.

Stringy instanton effects were studied in various contexts of string, M-theory and F-theory some time ago \cite{witten,ganor}. Recently, there was a revival of activity concerning phenomenological applications in D-brane models \cite{review}. These include neutrino masses \cite{blumencft1,blumencft2}, $\mu$-terms in the MSSM \cite{ibanez}, moduli stabilization \cite{cdmp},  supersymmetry breaking \cite{gaugemed} or fermion masses \cite{fermion}. An important feature of Euclidean brane instantons which contribute to superpotential couplings is that they have to wrap {\it rigid cycles}, i.e. they have no positions in the internal space. This requirement is necessary (although not sufficient) in order for the instanton to have the minimum number of (two) un-charged fermionic zero modes $\theta$ and to be therefore able to generate non-perturbative contributions to the superpotential.

\begin{figure}[!h]
\begin{center}
\includegraphics[width=8cm]{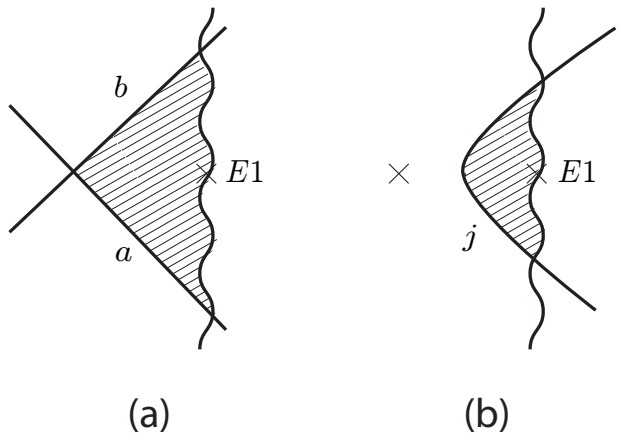}
\end{center}
\caption{\label{figw0} Non-perturbative Higgs mechanism through a
  dynamical condensate induced by a stringy Euclidean brane
  instanton. (a) Two fractional branes with opposite twisted charge
  are stuck in a given singularity. An instanton (represented by the
  wavy line) sits at a distant singularity. (b) The instanton may
  induce a condensate $\langle \Phi_{a\bar b}\tilde \Phi_{\bar a
    b}\rangle$ whose VEV parameterizes the position of the
recombined brane $j$ and minimizes the area of the disc amplitude spanned by the branes and the instanton (shaded region).}
\end{figure}

As their QFT analogs, stringy instanton effects can induce dynamical condensates and spontaneous breaking of symmetries in the 4d effective field theory. Consider for instance two stacks of fractional branes with opposite twisted charge stuck at an orbifold singularity. Their massless spectrum contains chiral multiplets $\Phi_{a\bar b}$ and $\tilde \Phi_{\bar a b}$ transforming, respectively, in the $(\textrm{Fund}(G_a),\overline{\textrm{Fund}(G_b)})$ and $(\overline{\textrm{Fund}(G_a)},\textrm{Fund}(G_b))$ representations of the gauge group $G_a\times G_b$. The condensate $\langle \Phi_{a\bar b}\tilde \Phi_{\bar a b}\rangle$ parameterizes the recombination of the two stacks of branes and the spontaneous breaking of the gauge group to a diagonal subgroup. The recombined stack of branes has vanishing total twisted charge and therefore can move off from the singularity, with position modulus related to $\Phi_{a\bar b}\tilde \Phi_{\bar a b}$. A non-trivial condensate $\langle \Phi_{a\bar b}\tilde \Phi_{\bar a b}\rangle$ can be induced by Euclidean brane instantons placed at distant singularities. The situation is depicted schematically in Figure \ref{figw0}. Varying $\langle \Phi_{a\bar b}\tilde \Phi_{\bar a b}\rangle$ changes the area of the disc scattering amplitude spanned by the branes and the instanton and hence the total energy of the system. Stringy instanton effects may therefore induce non-perturbative destabilization of a string vacuum and dynamical brane recombination.\footnote{See also \cite{blumencft1} for related comments.} Far away from the singularities, this phenomenon manifests in the appearance of a non-perturbative superpotential for the open string moduli of the recombined brane.

The purpose of the present work is to study this phenomenon in detail. We find, in particular, that if the following conditions are simultaneously satisfied:
\begin{itemize}
\item[-] there is an appropriate number of (two) charged instanton zero modes $\eta_i $ stretched between the brane and the instanton,
\item[-] there is an  $U(1)$ or $U(2)$ gauge theory realized on the D-branes,
\item[-] there is a Yukawa coupling $\eta_1 \Phi \eta_2 $, where $\Phi$ is a brane field charged under $U(1)$,
\end{itemize}
then there is a non-perturbatively generated linear superpotential $W_{n.p.} = g(\xi,\tau_i)e^{-S_{inst.}} \Phi$ in the gauge theory, where $S_{inst.}$ is the instanton action and the coefficient $g$ depends on the open string moduli $\xi$ and the complex structure moduli $\tau_i$. As we have just argued, the dependence on the open string moduli can be interpreted as a force acting on the D-brane by the singularity where the instanton resides, leading to an F-term contribution to the vacuum energy of the form $\Delta V \sim |g(\xi,\tau_i)e^{-S_{inst.}}|^2$.

In globally consistent models, there are usually several rigid instantons of action $S^{(\alpha)}_{inst.}$. The induced superpotential is then qualitatively of the form
\begin{equation}
 W_{n.p.} \ = \ \sum_{\alpha} g_{\alpha} ({\bf \xi}, \tau_i) e^{- S^{(\alpha)}_{inst.}} \Phi \ . \label{i1}
\end{equation}
The resulting scalar potential will be minimized for a certain position of the D-brane, which is typically a symmetric point with respect to the singularities where the various instantons sit. At these loci, non-perturbative forces are balanced.

In general, the functions $g_{\alpha}(\xi,\tau_i)$ contain also additional effects related to the dynamics of other D-branes which are needed for global consistency of the model. Even if these spectator branes do not experience any non-perturbative superpotential generation in their worldvolume, by moving towards the singularities where instantons sit they can change the number of instanton zero modes \cite{ganor} and therefore destroy the instability for the original D-branes under consideration. Therefore the notation ${\bf \xi}$ in eq.(\ref{i1}) should refer to the position of {\it all} D-branes and not only the ones experiencing the destabilizing force.

The generation of non-perturbative superpotentials for D-brane scalars
has many interesting applications. These range from moduli
stabilization to phenomenological applications such as models of
D-brane inflation \cite{inflation} or stringy realizations of
composite Higgs models. In addition, superpotentials of the form
eq.(\ref{i1}), which have an R-symmetry,  have also been recently employed to realize
supersymmetry breaking a la Polonyi and gauge mediation within string theory
\cite{gaugemed}. In this sense, it is generally accepted that if the number of complex fields on which $g_\alpha(\xi,\tau_i)$ depend is lower than the multiplicity of fields $\Phi$, the superpotential in eq.(\ref{i1}) spontaneously breaks  supersymmetry \cite{seiberg1,seiberg2}. This is certainly true when $g_\alpha(\xi,\tau_i)$ are arbitrary functions. However, our results suggest that in String Theory compactifications there are often correlations between the zeros of $g_\alpha(\xi,\tau_i)$, allowing for supersymmetric vacua even in cases where the non-perturbative superpotential preserves an R-symmetry. The reason is that due to the
particular dependence on $\xi$, the superpotential is not generic.

In this work we focus on toroidal orbifold models with magnetized branes and analyze the above process of non-perturbative vacuum destabilization, D-brane recombination and open string moduli stabilization. The requirement of having rigid cycles in models with magnetized branes implies the existence of a $\mathbb{Z}_2\times \mathbb{Z}_2$ subgroup within the orbifold action which has discrete torsion \cite{branesusybreak,cristinatorsion,blumenrigid,mikerigid}. One of our main results is the explicit expression for superpotential (\ref{i1}) in these models, provided the three conditions stated above are simultaneously satisfied.

The organization of the paper is as follows.  In Section \ref{ingredients} we summarize some of the ingredients which are needed for the computation of superpotentials like eq.(\ref{i1}), namely the definition of holomorphic fields in compactifications with magnetized D-branes, and the discrete parameters of stringy instantons in toroidal orbifold models. The explicit computation of superpotential (\ref{i1}) in toroidal orbifold models is performed in Section \ref{nonpert}, where we also comment on the general features of non-perturbative D-brane dynamics induced by stringy instantons and the limitations of the local analysis versus the global picture.
In Section \ref{simple} we present a simple global model based on the
$T^6/\mathbb{Z}_2\times \mathbb{Z}_2$ orbifold with magnetized branes
and rigid cycles, in which linear superpotential couplings can lead to
processes of D-brane recombination and gauge symmetry
breaking. We also briefly comment in this section on a different model with only magnetized D9-branes, where stringy instantons also generate linear superpotential couplings. We argue that in both cases supersymmetry is not broken.
Section \ref{globalmass} is devoted to the model presented
in Refs.\cite{cristinatorsion,mikerigid} where stringy instantons induce
mass terms for some of the antisymmetrics. By using these techniques we explicitly compute the mass terms, determining whether they
vanish or not. Finally, we end in Section \ref{conclusions} with some
conclusions, where we also comment about the implications for supersymmetry breaking. Three appendices collect some previous results on the $T^6/\mathbb{Z}_2\times \mathbb{Z}_2$ orbifold with magnetized branes and rigid cycles, various one-loop computations needed in the evaluation of the instanton one-loop determinants and details about the computation of F-terms in the global model of Section \ref{simple}.

\section{Some ingredients}
\label{ingredients}

In this Section we introduce some of the ingredients which are needed for the computation of superpotentials like eq.(\ref{i1}), namely the definition of holomorphic fields in compactifications with magnetized D-branes, and discrete configurations of Euclidean brane instantons in toroidal orbifold models. The computation of superpotential (\ref{i1}) will be then performed in Section \ref{nonpert}, by making use of these elements.

\subsection{Holomorphic fields in magnetized type I strings with Wilson lines}
\label{sechol}

In order to write down the effective superpotential (perturbative or non-perturbative) which results from a string theory compactification, one typically needs to compute some relevant physical couplings in the 4d action and to extract the holomorphic part. This procedure, however, is not in general straightforward. In the presence of D-branes, various non-holomorphic terms depending on the moduli of the D-branes are absorbed into redefinitions of the closed string moduli and charged chiral fields. Finding the precise form of these redefinitions is therefore required to correctly identify the holomorphic variables of the superpotential and, in a last step, the superpotential itself.

A systematic study of field redefinitions in the context of toroidal type IIB (orientifold) compactifications with magnetized D-branes and continuous Wilson lines was performed in \cite{ccd}. We reproduce here the main results of this paper, which will be relevant for later computations.

Hence, consider a stack of $N=\sum_I N_I$, $I=a,b,\ldots$, magnetized D9-branes wrapping a factorizable torus, $T^2\times T^2\times T^2$, and magnetization given by,
\begin{equation}
F_2=\sum_{r=1}^3\frac{\pi i}{\textrm{Im }\tau_r}\begin{pmatrix}\frac{m_a^r}{n_a^r}\mathbb{I}_{N_a}& & \\ & \frac{m_b^r}{n_b^r}\mathbb{I}_{N_b} & \\ & & \ddots \end{pmatrix}dz^r\wedge d\bar z^r
\end{equation}
with $n_I\in \mathbb{N}^+$, $m_I\in\mathbb{Z}$, such that $\textrm{g.c.d.}(m_I^r,n_I^r)=1$. Here, $\tau_r$ is the complex structure modulus of the $r$-th 2-torus, $dz^r=dx^r+\tau_r dy^r$, and $\mathbb{I}_{N_I}$ are $N_I \times N_I$ identity matrices. Magnetization breaks the initial $U(N)$ gauge symmetry to $\otimes_I U(N_I)$.

In addition, we consider configurations of continuous Wilson lines which do not further break the gauge symmetry, i.e.
\begin{equation}
A_{W.L.}=\sum_{r=1}^3\frac{\pi i}{\textrm{Im }\tau_r}\begin{pmatrix}\textrm{Im}(\xi_{a}^rd\bar z^r)\mathbb{I}_{N_a}& & \\
&\textrm{Im}(\xi_{b}^rd\bar z^r)\mathbb{I}_{N_b} & \\
& & \ddots  \end{pmatrix}
\end{equation}
where $\xi_{I,x}^r$, $\xi_{I,y}^r\in [0,1/n_I^r)$ are the real Wilson line parameters along the $r$-th 2-torus and $\xi_I^r = \xi_{I,x}^r + \tau_r \xi_{I,y}^r$.

In the presence of the above magnetization, $|I_{IJ}|$ chiral multiplets transforming in the bifundamental representation $(N_I,\bar N_J)$ of the gauge group arise in the massless spectrum, where $I_{IJ}$ is the intersection number\footnote{Although in this paper we work within type I string theory, very often we also refer to the T-dual type IIA terminology of intersecting branes.} between stacks of branes $I$ and $J$ defined as,
\begin{equation}
I_{IJ}=\prod_{r=1}^3I_{IJ}^r\ , \qquad I^r_{IJ}=m^r_I n^r_J - n^r_I m^r_J \ , \qquad r=1,2,3
\end{equation}
We denote the physical fields (also dubbed \emph{matter fields} in what follows) by $\Phi^{\vec i}_{IJ}$, where $\vec i =(i_1,i_2,i_3)$ and $i_r=0\ldots |I^r_{IJ}|-1$.

The $\mathcal{N}=1$ holomorphic variables which appear in the superpotential related to these fields can be extracted from the expression of the Yukawa couplings, resulting in \cite{ccd},\footnote{We omit in this expression factors depending on the K\"ahler moduli or the axion-dilaton. These terms will not be relevant for the computations in this paper.}
\begin{equation}
\hat\Phi_{ab}^{\vec j}=e^{if_{ab}}\left(\frac{ W}{\overline W}\right)^{\frac14}\left(\prod_{r=1}^3(\textrm{Im }\tau_r)^{\frac14}e^{i\pi\frac{\xi^r_{ab}\textrm{Im }\xi^r_{ab}}{\mathcal{I}^{r}_{ab}\textrm{Im }\tau_r}}\right)\Phi_{ab}^{\vec j}\label{var}
\end{equation}
where we have introduced the notation
\begin{equation}
\mathcal{I}_{ab}^r = {I_{ab}^r\over n_an_b}\ , \qquad
\xi_{ab}^r=\xi_a^r-\xi_b^r
\end{equation}
The phases $f_{ab}$ are unknown real functions of the Wilson line moduli. By demanding invariance of the 4d action under periodic shifts of the Wilson line moduli, however, their transformation properties can be derived. More precisely, the holomorphic variable (\ref{var}) has to transform as
\begin{align}
\xi_a^r &\rightarrow \xi_a^r + \delta_a^r  \quad\quad \Rightarrow  \quad  \hat\Phi^{\vec i}_{ab}\ \to \ \hat\Phi^{\vec i}_{ab}\ , \label{hol2a}\\
\xi_a^r &\rightarrow \xi_a^r + \delta_a^r \tau_r \quad \Rightarrow \quad \hat\Phi^{\vec i}_{ab}\ \to \ e^{\frac{i\pi(\delta^r_a)^2\tau_r}{\mathcal{I}^r_{ab}}+\frac{2\pi i\xi_{ab}^r\delta_a^r}{\mathcal{I}^r_{ab}}}\hat\Phi^{\vec i}_{ab}\
\label{hol2}
\end{align}
where $\delta_a^r = \textrm{l.c.m.}(I_{ab}^r,I_{ac}^r,\dots)/n_a^r$. From here, we deduce that the phase factors $f_{ab}=-f_{ba}$ transform as,
\begin{align}
\xi_a^r &\rightarrow \xi_a^r + \delta_a^r  \quad\quad \Rightarrow  \quad  f_{ab}\ \to \ f_{ab}-\frac{\pi \delta_a^r\textrm{Im }\xi_{ab}^r}{\mathcal{I}_{ab}^r\textrm{Im }\tau_r}\ , \label{trasf1}\\
\xi_a^r &\rightarrow \xi_a^r + \delta_a^r \tau_r \quad \Rightarrow \quad f_{ab}\ \to \ f_{ab}-\frac{\pi \delta_a^r\textrm{Im}(\bar\tau_r\xi_{ab}^r)}{\mathcal{I}_{ab}^r\textrm{Im }\tau_r}\
\label{trasf2}
\end{align}

Closed string moduli are also redefined by the open string moduli (see e.g. \cite{bachas,otrosbachas,berg} for early examples) absorbing non-holomorphic terms in their definition. Redefinitions of the complex axion-dilaton $S$ and the K\"ahler moduli $T_k$ can be deduced by looking at non-perturbative couplings generated by $E5$ and $E1$-brane instantons \cite{ccd}. For factorizable 6-tori, these turn out to be given by,
\begin{align}
\hat S & = S\ +\ \sum_{\{ I\}}\sum_{r=1}^3c_{ I}^r\frac{\xi^r_{I}\textrm{Im }\xi^r_{ I}}{\textrm{Im }\tau_r}\label{ss}\\
\hat T_k &= T_k\ -\ \sum_{\{ I\}}\left[c_{ I}^0\frac{\xi^k_{ I}\textrm{Im }\xi^k_{ I}}{\textrm{Im }\tau_k}-\sum_{p\neq q\neq k}^3c_{ I}^q\frac{\xi^p_{ I}\textrm{Im }\xi^p_{ I}}{\textrm{Im }\tau_p}\right]\ , \qquad k=1,2,3\label{tt}
\end{align}
where $S$ and $T_k$ obey the standard definitions for factorizable toroidal orientifold compactifications with O9-planes \cite{bachas,luis,effec},
\begin{equation}
S = C_6+ie^{\phi/2}\prod_{r=1}^3\textrm{Vol}_{r}\ , \qquad T_k = C_{2,k}+ie^{-\phi/2}\textrm{Vol}_{k}
\end{equation}
In these expressions, $\phi$ is the 10d dilaton, $\textrm{Vol}_{r}$ is the volume of the $r$-th 2-torus and $C_6$ ($C_{2,k}$) is the component of the RR 6-form (2-form) along the compact space (the $k$-th 2-torus).

The index $I$ in eqs.(\ref{ss})-(\ref{tt}) runs over all D-branes
in the compactification, magnetized and non-magnetized, whereas the
coefficients $c^0_{I}$ and $c^k_{I}$, $k=1,2,3$ correspond,
respectively, to the D9- and D5-brane charges of the stack of branes $\hat I$. For magnetized bulk D9-branes these are given by,
\begin{equation}
c_{I}^0=N_{I} n_{I}^1n_{I}^2n_{I}^3\ , \quad
c_{I}^1=N_{I} n_{I}^1m_{I}^2m_{I}^3\ , \quad
c_{I}^2=N_{I} m_{I}^1n_{I}^2m_{I}^3\ , \quad
c_{I}^3=N_{I} m_{I}^1m_{I}^2n_{I}^3\ .
\end{equation}
Alternatively, their values can be read from the mixing between different closed string moduli in the gauge kinetic function, $f_{I}$.

Any consistent superpotential should be a holomorphic function of the variables (\ref{ss})-(\ref{tt}). However, in order for (\ref{ss})-(\ref{tt}) to transform holomorphically under periodic shifts of the Wilson line moduli, the RR potentials have to transform also non-trivially. From the 10d supergravity action it can be seen that this is indeed the case. For instance, integrating over the internal 6-torus the  following piece of the 10d supergravity action
\begin{equation}
\int\left[\textrm{Tr}(F_2\wedge F_2\wedge F_2\wedge A)\wedge F_3+F_7\wedge F_3\right]
\end{equation}
where $F_3$ and $F_7$ are respectively the RR 3-form and 7-form field strengths ($F_7=*F_3$), one obtains
\begin{equation}
\frac16\int dx^4 \ \epsilon^{\mu\nu\rho\sigma}F_{\nu\rho\sigma}\left[\textrm{Re}(\partial_\mu S)+
\sum_{I}\left( N_{I} m^1_I m^2_I m^3_I A_\mu^I +\sum_{r=1}^3c_I^r\frac{\textrm{Im}[\bar \xi^r_I\partial_\mu\xi^r_I]}{\textrm{Im }\tau_r}\right)\right]\label{gs}
\end{equation}
The first two terms in the bracket are the ones responsible for the Green-Schwarz mechanism. From them we see that a $U(1)$ gauge transformation has to be compensated by a shift of the closed string axion,
\begin{equation}
A^I_\mu \to A^I_\mu + \partial_\mu\Lambda^I \quad \Rightarrow \quad S\to
S - N_{I} m^1_I m^2_I m^3_I\Lambda^I
\end{equation}
Similarly, the third term leads to non-trivial transformations of $S$ under periodic shifts of the Wilson line moduli,
\begin{align}
\xi^r_a&\to\xi^r_a+\delta^r_a \  \quad\quad \Rightarrow \quad
S  \to  S  - c^r_a\frac{\delta_a^r\textrm{Im }\xi^r_a}{\textrm{Im }\tau_r}\label{shift1}\\
\xi^r_a&\to\xi^r_a+\delta^r_a \tau_r\ \quad \Rightarrow \quad
S   \to  S  - c^r_a\frac{\delta_a^r\textrm{Im }(\xi^r_a\bar\tau_r)}{\textrm{Im }\tau_r}\label{shift2}
\end{align}
Plugging these equations into eq.(\ref{ss}) we observe that $\hat S$ indeed transforms holomorphically,
\begin{align}
\xi^r_a&\to\xi^r_a+\delta^r_a \  \quad\quad \Rightarrow \quad \hat S  \to \hat S\\
\xi^r_a&\to\xi^r_a+\delta^r_a\tau_r \ \quad  \Rightarrow \quad \hat S  \to \hat S +2c_a^r\delta^r_a\xi_a^r+(\delta_a^r)^2c_a^r\tau_r
\end{align}
providing a good consistency check of eq.(\ref{ss}).

Similar considerations apply to $T_k$. In this case, under $U(1)$ gauge transformations
\begin{equation}
A^I_\mu \to A^I_\mu + \partial_\mu \Lambda^I \quad \Rightarrow \quad
T_k\to T_k + N_{I} n^p_I n^q_I m^r_I \Lambda^I \ , \qquad p\neq q\neq k\label{trasg}
\end{equation}
whereas under periodic shifts of the Wilson line moduli,
\begin{align}
\xi^r_a&\to\xi^r_a+\delta^r_a \ \quad\quad \Rightarrow \quad
T_k \to T_k +c^{r,k}_a\frac{\delta_a^r\textrm{Im }\xi_a^r}{\textrm{Im }\tau_r}\label{trasa1}\\
\xi^r_a&\to\xi^r_a+\delta^r_a\tau_r \ \quad \Rightarrow \quad
T_k \to T_k +c^{r,k}_a\frac{\delta_a^r\textrm{Im}(\xi_a^r\bar\tau_r)}{\textrm{Im }\tau_r}\label{trasa2}
\end{align}
where we have defined $c^{r,k}_a=c^0_a$ for $r=k$ and $c^{r,k}_a=-c^p_a$ for $r\neq k\neq p$. Plugging these expressions into eq.(\ref{tt}) leads to the holomorphic transformation of $\hat T_k$,
\begin{align}
\xi^r_a&\to\xi^r_a+\delta^r_a \  \quad\quad \Rightarrow \quad
\hat T_k \to \hat T_k\, \label{tras1}\\
\xi^r_a&\to\xi^r_a+\delta^r_a\tau_r \ \quad \Rightarrow \quad
\hat T_k \to \hat T_k -2c^{r,k}_a\delta^r_a\xi_a^r-(\delta_a^r)^2c^{r,k}_a\tau_r\label{tras2}
\end{align}

In this work we will consider $\mathcal{N}=1$ compactifications of type I
string theory on $T^6/\Gamma$, where $\Gamma\subset SU(3)$ is some
discrete orbifold action. The above results hold in this case as long
as we take them in the covering space, in the same spirit than \cite{bkors}.
For instance, consider a field which transforms in an antisymmetric or
symmetric representation of the gauge group of
a stack of branes $a$. Such a field would result from open strings in
the $a-a^*$ sector, where $a^*$ is the image of $a$
under the orientifold action. Wilson line moduli are mapped as
$\xi^r_{a^*}=-\xi^r_{a}$ \cite{iiayukawa}. Hence, when performing a shift of the modulus $\xi^r_a$, one has also to shift $\xi^r_{a^*}$ conveniently. From eqs.(\ref{hol2a})-(\ref{hol2}) we obtain then the holomorphic transformation of such a field,
\begin{align}
\xi^r_a&\to\xi^r_a+\delta^r_a \  \quad\quad \Rightarrow \quad \hat\Phi^{\vec i}_{aa^*}\ \to \ \hat\Phi^{\vec i}_{aa^*}\, \nonumber\\
\xi^r_a&\to\xi^r_a+\delta^r_a \tau_r\ \quad \Rightarrow \quad \hat\Phi^{\vec i}_{aa^*}\ \to \ e^{\frac{4\pi i(\delta^r_a)^2\tau_r}{\mathcal{I}^r_{aa^*}}+\frac{8\pi i\xi_{a}^r\delta_a^r}{\mathcal{I}^r_{aa^*}}}\hat\Phi^{\vec i}_{aa^*} \nonumber
\end{align}

\subsection{Stringy instantons in toroidal orbifold models}
\label{stringy}

In general, the 4d perturbative effective
action of a type I string theory compactification receives corrections from E1 and E5-branes
wrapping, respectively, holomorphic complex curves or the whole
compact space. We are mainly interested on corrections to the
superpotential. These are due to Euclidean branes with precisely two
fermionic neutral zero modes, only charged under the gauge group of
the instanton \cite{review}. In terms of the 4d effective
supergravity, these are identified with the
fermionic coordinates of the $\mathcal{N}=1$ superspace. Additional
fermionic neutral zero modes would render the contribution to the
superpotential trivial, as they lead to vanishing Grassman
integrals over the instanton moduli space.\footnote{Such instantons
  could however contribute to
other quantities in the 4d effective theory, such as the gauge kinetic
function \cite{cdmp,max,cd1,lerda}, the K\"ahler potential \cite{cd1}
or higher order F-terms \cite{higher,higher0,higher1,higher2,higher3,higher4}.} Notice however that in some cases these extra
zero modes can be removed by interactions between instanton zero
modes \cite{higher1,lift2,petersson} or/and additional fluxes \cite{higher1,flux,higher3}.

Geometrically, the condition of having only two fermionic neutral modes in the massless spectrum, among other things, requires the cycle wrapped by the instanton to be rigid. For toroidal orbifold models with magnetized branes this can only be the case if the orbifold group $\Gamma$ contains a $\mathbb{Z}_2\times \mathbb{Z}_2$ subgroup with discrete torsion \cite{branesusybreak,cristinatorsion,blumenrigid,mikerigid}. We denote the elements of this subgroup as $g$, $f$ and $h$, where each generator reverses the coordinates of two 2-tori within the $T^6$,
\begin{equation}
g:(+,-,-) \ , \qquad f:(-,+,-) \ , \qquad h:(-,-,+) \ .\label{orbifold}
\end{equation}
Thus, in total there are $3\times 16$ fixed points of the orbifold action associated to this $\mathbb{Z}_2\times \mathbb{Z}_2$ subgroup.

The sign choice for the disconnected modular orbits of $g$, $f$ and $h$ twisted sectors in the one-loop partition function leads to discrete torsion. The later is usually parameterized in terms of three signs $(\epsilon_g,\epsilon_f,\epsilon_h)$. From the point of view of the target space, the choice of discrete torsion determines the nature of the exceptional divisors at the above $3\times 16$ singularities and their parity under the orientifold involution. For $\epsilon_g\epsilon_f \epsilon_h=+$ each of the fixed points leads to a (1,2)-cycle in the blown-up limit, whereas for $\epsilon_g\epsilon_f \epsilon_h=-$ it leads to a (1,1)-cycle.
For the sake of concreteness, here we set
$(\epsilon_g,\epsilon_f,\epsilon_h)=(+,+,-)$. This requires the
presence of exotic O$5_+$-planes along the 2-torus invariant under $h$
\cite{branesusybreak}. Other cases easily follow from this one.

Fractional E1-branes wrapping the third 2-torus and stuck at the 16
fixed points of $h$ have precisely two fermionic neutral zero modes,
and therefore may correct the superpotential. Their wrapping numbers
are
\begin{equation}
(m^i , n^i) \ = \ (1 , 0 ) \otimes (-1 , 0 ) \otimes (0 , 1 ) \ .
\end{equation}
We choose to parameterize the positions of the $16$ fixed points and the configuration of discrete Wilson lines in the worldvolume of E1-branes by means of three discrete complex parameters,
\begin{equation}
\xi^i_{\rm E1} = \epsilon^i + \tau_i \epsilon^{i+3} \ , \qquad i=1,2,3\label{E1pos}
\end{equation}
with $\epsilon^i$ taking values $0$ or $1/2$. Here, $i=1,2,3$ denote the 2-tori invariant under the action of $g$, $f$ and $h$, respectively. Notice that, contrary to the case of standard D$p$-branes, $\xi^i_{\rm E1}$ are not physical fields. The instanton action lacks the corresponding kinetic terms, since the instanton does not span the non-compact space-time dimensions, and therefore these fields do not propagate.

In general, the presence of D$p$-branes intersecting the instanton gives rise to extra fermionic zero modes charged under the diagonal $U(1)$ gauge symmetry of the D$p$-brane.
The integral over these fermionic zero modes has to be saturated by gauge invariant operators, where each zero mode appears exactly once. Gauge invariance generically requires these operators to contain also insertions of matter fields, $\Phi_q$, localized at the intersections between stacks of D$p$-branes \cite{review}. After integrating over the moduli space of the instanton, a non-perturbative superpotential coupling for the matter fields $\Phi_q$ is induced.

\section{Non-perturbative D-brane dynamics}
\label{nonpert}

\subsection{Linear term instabilities}

Following our discussion in the introduction, we are interested in compactifications where Euclidean brane
instantons generate linear superpotential couplings for some of the
charged matter fields. Those couplings would
signal a non-perturbative instability in the initial configuration, or
in the most favorable case, the spontaneous breaking of the
$\mathcal{N}=1$ supersymmetry and, eventually, gauge mediation transmission.

Hence, consider a Yukawa coupling in the instanton action involving a chiral field, $\Phi_{ab}$, which transforms in the bifundamental representation of $U(1)_a\times U(1)_b$, and two instanton charged fermionic zero modes, $\eta_a$ and $\eta_b$, charged under each $U(1)$ factor respectively,
\begin{equation}
S_{\textrm{inst.}}=S_{E1}+g(\xi,\tau^k)\eta_a\Phi_{ab}\eta_b\label{start}
\end{equation}
In this expression, $S_{E1}$ is the part of the tree level instanton action which only involves moduli. In the particular case at hand it is given by,
\begin{equation}
S_{E1}=T_3\label{treee}
\end{equation}
where we have set the VEV of blow-up moduli to zero.\footnote{In the most general case, each term in an
instantonic sum would be weighted by an exponential factor depending on the particular linear combination of blow-up moduli which the instanton couples to.} Note that, according to the discussion in Section \ref{sechol}, $S_{E1}$ shifts non-trivially under both, $U(1)$ gauge transformations and periodic shifts of the Wilson line moduli.

After integration of eq.(\ref{start}) over the instanton charged zero modes, the following non-perturbative contribution to the effective superpotential results \cite{mikerigid},
\begin{equation}
W_{n.p.}=g(\xi,\tau^k)e^{ 2 \pi i S_{E1}}\Phi_{ab}\label{npsuper}
\end{equation}
The zeroes of $g(\xi,\tau^k)$ will therefore determine the non-perturbative supersymmetric vacuum.

Similarly, if the two $U(1)$ factors are identified by some discrete orbifold or orientifold symmetry,
such that the massless spectrum contains chiral fields $A_{ij}$ ($S_{12}$) transforming in the antisymmetric (symmetric) representation of $U(2)$ ($U(1)$) and one (two) instanton charged fermionic zero modes,
\begin{align}
S_{\textrm{inst.}}&=S_{E1}+g(\xi,\tau^k)\sum_{i,j=1}^2\eta_iA_{ij}\eta_j &\Rightarrow \qquad W_{n.p.}&=g(\xi,\tau^k)e^{ 2 \pi i S_{E1}}\sum_{i,j=1}^2\epsilon_{ij}A_{ij}\\
S_{\textrm{inst.}}&=S_{E1}+g(\xi,\tau^k)\eta_1S_{12}\eta_2 &\Rightarrow \qquad W_{n.p.}&=g(\xi,\tau^k)e^{ 2 \pi i S_{E1}}S_{12}
\end{align}

Since toroidal orbifold models admit a description in terms of a CFTs, we can make use of the D-brane instanton calculus \cite{blumencft1,blumencft2} to
explicitly compute the above non-perturbative F-terms. Diagrammatically, the physical F-term associated to a chiral field $\Phi_{ab}$ which results from open strings stretched between two stacks of magnetized branes $a$ and $b$ can be expressed as,
\begin{center}
\includegraphics[width=13.5cm]{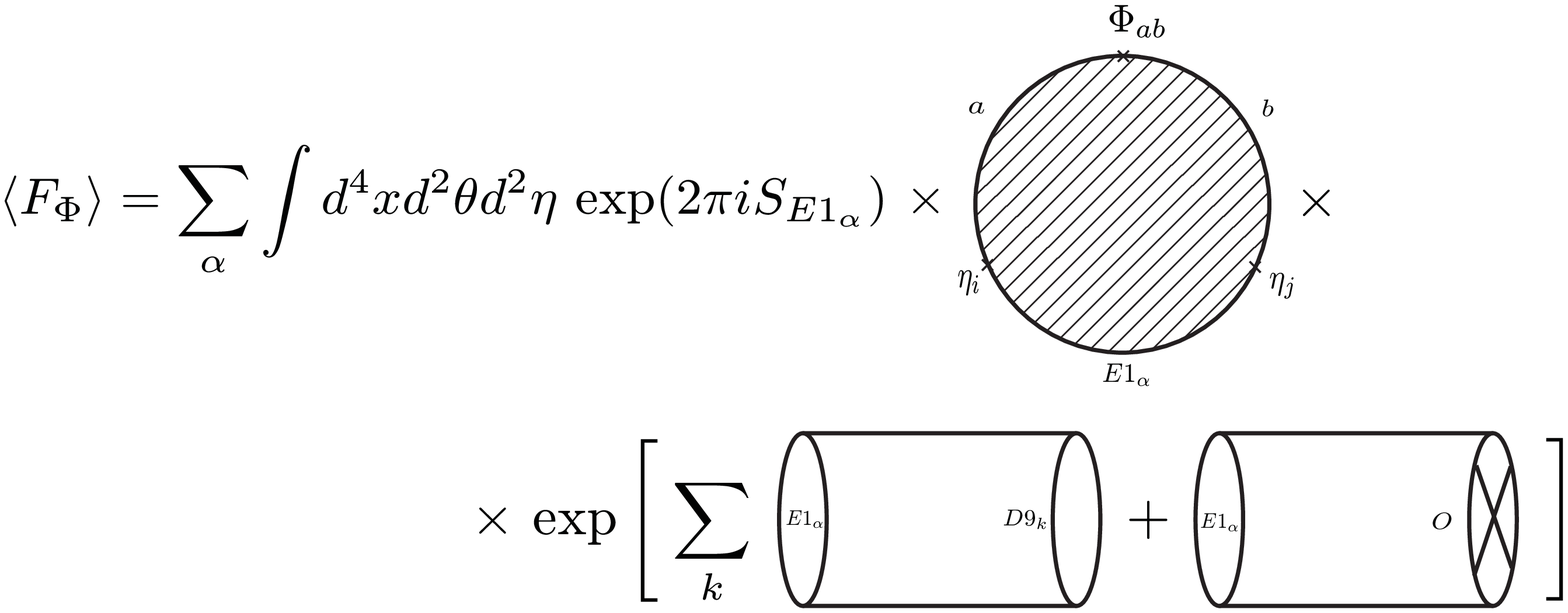}
\end{center}
where the sum in $\alpha$ extends over all possible configurations for the position, discrete Wilson lines and Chan-Paton charge of the instanton. In general, there are $|I_{ab}|$ different F-terms of this form, corresponding  to the chiral fields $\Phi_{ab}^{\vec i}$, $i_r=0\ldots |I^r_{ab}|-1$. Note that
the F-auxiliary field of these chiral
multiplets carries a $\theta^2$ term, $\Phi_{ab}=\phi+F_\Phi\theta^2$,
so the integral over the instanton fermionic zero modes is non-vanishing. F-terms corresponding to fields $A_{ij}$ or $S_{12}$ equally follow from $\langle F_\Phi\rangle$, by modding out its expression with the appropriate discrete symmetry.

Both, disc and annulus scattering amplitudes, contain
arbitrary insertions of open string moduli,
parameterizing the energy change induced by deformations of the
D-brane sector. Thus,  $\langle F_\Phi\rangle$
is in general a complicated function of the open string moduli. The disc scattering amplitude is related to the
shaded area in Figure \ref{figw0},
responsible for the destabilizing force. On the other hand, annulus
and M\"obius scattering amplitudes stretch between
the instanton and other D-branes and O-planes present in the
compactification. From the point of view of the 4d effective theory,
these are related to one-loop determinants of fluctuations around
the instanton background \cite{review}.

Let us compute $\langle F_\Phi\rangle$ explicitly. Yukawa couplings for toroidal compactifications with magnetized D-branes have been computed in \cite{yukawa} by dimensionally
reducing 10d super Yang-Mills theory (see also
\cite{iiayukawa,divecchia,antoniadis,choi}). The 3-point disc scattering
amplitude involving two magnetized branes and a
rigid E$1$-brane with two fermionic charged instanton zero modes takes the form,
\begin{equation}
\mathcal{D}_{abE1}^{\vec i}\sim e^{i (f_{ab}+f_{bE1}+f_{E1a})}\prod_{r=1}^3\textrm{exp}\left[\frac{\pi i
      \xi^r_{abE1} \textrm{Im
      }\xi^r_{abE1}}{|I^r_{ab}|\textrm{Im }\tau_r}\right]
\vartheta\left[{\frac{i_r}{I^r_{ab}} \atop 0}\right](\xi^r_{abE1}\ ;\ \tau_r|I^r_{ab}|) \label{yuk}
\end{equation}
We have omitted in this expression normalization factors
coming from the K\"ahler metric since they are irrelevant for our purposes.

Continuous Wilson line moduli of branes $a$ and $b$ along the $r$-th 2-torus, $\xi^r_I$, and discrete parameters $\xi^r_{E1}$ of the instanton, eq.(\ref{E1pos}), are all combined in the quantity,
\begin{equation}
\xi^r_{abE1}=I_{ab}^r\xi_{\rm E1}^{r}+I^r_{E1a}\xi^{r}_{b}+I^r_{bE1}\xi^{r}_a\ ,
\end{equation}
with $I^r_{E1a}\ (I^r_{bE1})=\pm 1$ denoting the chirality of the charged zero mode localized at the $a-E1$ ($E1-b$) intersection. Moreover, since there is no exchange of twisted fields at the disc level, the amplitude is independent of the Chan-Paton charges of the branes and the instanton.

The contribution of this scattering amplitude to the superpotential can be easily extracted once the holomorphic $\mathcal{N}=1$ variables are identified. For that, we make use of the results in Section \ref{sechol}. In order $W\sim\langle F_\Phi\rangle\Phi_{ab}$ to be invariant under $U(1)$ gauge transformations and periodic shifts of the Wilson line moduli, the integration variables $\eta_a$ and $\eta_b$ appearing in $\langle F_\Phi\rangle$ should transform in the same way than the term exp$(2\pi i S_{E1})$. Thus, looking at eqs.(\ref{trasa1})-(\ref{trasa2}) and (\ref{treee}), we observe that under $U(1)$ gauge transformations,
\begin{align}
A_\mu^a &\to A_\mu^a + \partial_\mu \Lambda^a \quad \Rightarrow \quad \eta_a \to e^{2\pi i \Lambda^a} \ \eta_a\ , \quad \eta_b \to \eta_b\ , \label{jk1}\\
A_\mu^b &\to A_\mu^b + \partial_\mu \Lambda^b \quad \Rightarrow \quad \eta_a \to \eta_a\ , \quad \eta_b \to e^{-2\pi i \Lambda^b} \ \eta_b\ ,
\end{align}
whereas under shifts of the Wilson line moduli,
\begin{align}
\xi^r_a & \to \xi^r_a+\delta_a^r \quad \Rightarrow \quad \eta_a \to e^{-\frac{\pi i \delta_a^r I^r_{E1a}\textrm{Im }\xi^r_a}{\textrm{Im }\tau_r}}\eta_a\ , \quad \eta_b\to\eta_b\ , \\
\xi^r_b & \to \xi^r_b+\delta_b^r \quad \Rightarrow \quad \eta_a \to \eta_a\ , \quad \eta_b\to e^{-\frac{\pi i \delta_b^r I^r_{bE1}\textrm{Im }\xi^r_b}{\textrm{Im }\tau_r}}\eta_b\ ,
\end{align}
and,
\begin{align}
\xi^r_a & \to \xi^r_a+\delta_a^r\tau_r \quad \Rightarrow \quad \eta_a \to e^{-\frac{\pi i \delta_a^r I^r_{E1a}\textrm{Im}(\xi^r_a\bar\tau_r)}{\textrm{Im }\tau_r}}\eta_a\ , \quad \eta_b\to\eta_b\ , \\
\xi^r_b & \to \xi^r_b+\delta_b^r\tau_r \quad \Rightarrow \quad \eta_a \to \eta_a\ , \quad \eta_b\to e^{-\frac{\pi i \delta_b^r I^r_{bE1}\textrm{Im}(\xi^r_b\bar\tau_r)}{\textrm{Im }\tau_r}}\eta_b\ ,\label{jk2}
\end{align}
with $\delta^r_a=\textrm{l.c.m.}(I^r_{ab},I^r_{ac},\ldots)/n^r_a$. These transformations allow us to fix the correct normalization for the instanton fermionic charged zero modes (remember that there are not kinetic terms for these modes in the instanton action). Note, in particular, that eq.(\ref{yuk}) is invariant under periodic shifts of the Wilson line moduli $\xi^r_K$, reflecting the fact that it assumes a normalization for the charged zero modes such that they are charged under U(1) gauge transformations but are invariant under shifts of the Wilson line moduli. Let us denote these ``wrong'' variables as $\lambda_a$ and $\lambda_b$. Making use of the following identity,
\begin{multline}
\textrm{exp}\left[\frac{i\pi
      \xi^r_{abE1} \textrm{Im
      }\xi^r_{abE1}}{|I^r_{ab}|\textrm{Im }\tau_r}\right]
\vartheta\left[{\frac{i_r}{I^r_{ab}} \atop 0}\right](\xi^r_{abE1}\ ;\ \tau_r|I^r_{ab}|) = \\
\exp \left[ i\pi\left(\frac{(I^r_{bE1}\xi^r_a+I^r_{E1a}\xi^r_b)\textrm{Im}(I^r_{bE1}\xi^r_a+I^r_{E1a}\xi^r_b)}{|I^r_{ab}|\textrm{Im }\tau_r}+I^r_{bE1}\frac{\textrm{Im}(\xi^r_b\bar\xi^r_{E1})}{\textrm{Im }\tau_r}+I^r_{E1a}\frac{\textrm{Im}(\xi_a^r\bar\xi_{E1}^r)}{\textrm{Im }\tau_r}\right)\right] \\
\times e^{-i\pi\epsilon^r\epsilon^{r+3}|I^r_{ab}|}\vartheta\left[{\frac{i_r}{I^r_{ab}}+ \epsilon^{r+3}
    \atop  I_{ab}^r \epsilon^{r} }\right] ( I^r_{bE1}\xi_a^r+I^r_{E1a}\xi^r_b\ ;\
\tau_r|I^r_{ab}|) \nonumber
\end{multline}
we can then define,
\begin{align}
\eta_a & = \textrm{exp}\left[i\left(f_{E1a}+\pi\sum_{r=1}^3I^r_{E1a}\frac{\textrm{Im}(\xi_a^r\bar\xi_{E1}^r)}{\textrm{Im }\tau_r}\right)\right]\ \lambda_a\label{eta1}\\
\eta_b & = \textrm{exp}\left[i\left(f_{bE1}+\pi\sum_{r=1}^3I^r_{bE1}\frac{\textrm{Im}(\xi_b^r\bar\xi_{E1}^r)}{\textrm{Im }\tau_r}\right)\right]\ \lambda_b\label{eta2}
\end{align}
so that, taking into account the transformation properties of phase factors $f_{IJ}$ given in eqs.(\ref{trasf1})-(\ref{trasf2}), one may verify that $\eta_a$ and $\eta_b$ defined in (\ref{eta1}) and (\ref{eta2}) indeed transform according to (\ref{jk1})-(\ref{jk2}) under $U(1)$ gauge transformations and shifts of the Wilson line moduli.

We have thus to evaluate,
\begin{multline}
W_{\rm disc} =\int d\eta_a d\eta_b \ \eta_a\eta_b\Phi^{\vec i}_{ab}\ \prod_{r=1}^3 \sum_{\epsilon^r,\epsilon^{r+3}=0,\frac12} \vartheta\left[{\frac{i_r}{I^r_{ab}}+ \epsilon^{r+3}
    \atop  I_{ab}^r \epsilon^{r} }\right] ( I^r_{bE1}\xi_a^r+I^r_{E1a}\xi^r_b\ ;\
\tau_r|I^r_{ab}|) \\
\exp \left[2\pi i T_3+if_{ab} + i\pi\frac{(I^r_{bE1}\xi^r_a+I^r_{E1a}\xi^r_b)\textrm{Im}(I^r_{bE1}\xi^r_a+I^r_{E1a}\xi^r_b)}{|I^r_{ab}|\textrm{Im }\tau_r}-i\pi\epsilon^r\epsilon^{r+3}|I^r_{ab}|\right]
\end{multline}
which, in terms of the holomorphic variables, eqs.(\ref{var}) and (\ref{tt}), leads to,
\begin{multline}
W_{\rm disc} = \\ \prod_{r=1}^3 \sum_{\epsilon^r,\epsilon^{r+3}=0,\frac12} e^{2 \pi i \hat T_3- i \pi |I^r_{ab}| \epsilon^r
  \epsilon^{r+3}} \vartheta\left[{\frac{i_r}{I^r_{ab}}+  \epsilon^{r+3}
    \atop  I_{ab}^r \epsilon^{r} }\right] (I^r_{bE1}\xi_a^r+I^r_{E1a}\xi^r_b\ ;\
\tau^r|I^r_{ab}|)  \ \hat \Phi_{ab}^{\vec i}  \label{3.2.3}
\end{multline}
This is an holomorphic function of the $\mathcal{N}=1$ chiral variables associated to the physical fields, periodic under shifts of the Wilson line moduli and $U(1)$ gauge transformations (c.f. eqs.(\ref{tras1})-(\ref{tras2}) and (\ref{trasg})), providing strong checks of its consistency.

Next, let us evaluate the contributions to $\langle F_\Phi\rangle$ coming from
one-loop string scattering amplitudes. One may verify that annulus amplitudes stretching between the
instanton and magnetized (bulk or fractional) branes are independent
of the open string moduli and only correct normalization factors in eq.(\ref{yuk}).
The computation of the remaining
one-loop M\"obius and annulus amplitudes is summarized in Appendix \ref{thresholds} for $T^6/\mathbb{Z}_2\times\mathbb{Z}_2$, leading to the expression,\footnote{Similar results can be
obtained from the computation of the gauge threshold corrections for
non-magnetized D5-branes \cite{berg,maldacena,lust}.}
\begin{multline}
\textrm{exp}\left(\mathcal{A}_{E1}+\mathcal{M}_{E1}\right)
\sim \left[\eta(\tau_1)^{1+2N_{D5_2}}
\eta(\tau_2)^{1+2N_{D5_1}}\eta(\tau_3)^{1+2N_{D9}}\right]^{-1}\times\\
\times\prod_{K=1}^{N_{D9}}\left(\textrm{exp}\left[\frac{2\pi i (\xi^3_{K}
      \textrm{Im }\xi^3_{K} +\xi^3_{E1} \textrm{Im
      }\xi^3_{E1})}{\textrm{Im }\tau_3}\right]
\vartheta\left[{\frac{1}{2} \atop \frac{1}{2}}\right](\xi^3_K
+\xi^3_{E1} ; \tau_3)  \vartheta\left[{\frac{1}{2} \atop \frac{1}{2}}\right](\xi^3_K
-\xi^3_{E1} ; \tau_3) \right)  \\
\times\prod_{Q=1}^{N_{D5_1}}
\left(\textrm{exp}\left[\frac{2\pi i (\xi^2_{Q}
      \textrm{Im }\xi^2_{Q} +\xi^2_{E1} \textrm{Im
      }\xi^2_{E1})}{\textrm{Im }\tau_2}\right]
\vartheta\left[{\frac{1}{2} \atop \frac{1}{2}}\right](\xi^2_Q
+\xi^2_{E1} ; \tau_2)  \vartheta\left[{\frac{1}{2} \atop
    \frac{1}{2}}\right] (\xi^2_Q -\xi^2_{E1} ; \tau_2) \right)\\
\times\prod_{P=1}^{N_{D5_2}}
\left(\textrm{exp}\left[\frac{2\pi i (\xi^1_{P}
      \textrm{Im }\xi^1_{P} +\xi^1_{E1} \textrm{Im
      }\xi^1_{E1})}{\textrm{Im }\tau_1}\right]
\vartheta\left[{\frac{1}{2} \atop \frac{1}{2}}\right](\xi^1_P
+\xi^1_{E1} ; \tau_1)  \vartheta\left[{\frac{1}{2} \atop
    \frac{1}{2}}\right] (\xi^1_P -\xi^1_{E1} ; \tau_1) \right)\nonumber
\end{multline}
where we have considered the contribution of $N_{D9}$ non-magnetized
bulk D9-branes and $N_{D5_s}$ D$5$-branes, $s=1,2$, wrapping the $s$-th 2-torus (also dubbed D$5_s$-branes in what follows). According to this,  $\xi^3_K$ is the complexified continuous Wilson line modulus of the $K$-th bulk D9-brane along the third 2-torus, and $\xi^{1}_P$ ($\xi^{2}_Q$) is
the complexified position modulus of the $P$-th D$5_2$-brane (the $Q$-th D$5_1$-brane) along
the first (second) 2-torus. Note that in general we can distinguish two types of bulk branes, depending on whether they are charged under one of the twisted sectors or they carry no twisted charge at all. In the first case the brane can only move along a single 2-torus (the one which is left invariant by the corresponding orbifold generator), whereas in the second case it can move along the full compact space. In this regard, we define by convention the number of branes to be given by the rank of the corresponding gauge group. For instance, $N_{D9}=1$ denotes a bulk D9-brane with twisted charge under $h$, so that it can only move along the third torus, whereas $N_{D9}=2$ denotes a bulk D9-brane with no twisted charge of any twisted sector. Analogous definitions apply to bulk D$5$-branes.

The one-loop contribution to the holomorphic superpotential computed from this expression then reads \cite{ccd}
\begin{multline}
W_{\rm one-loop} =\\
\left[\eta(\tau_1)^{1+2N_{D5_2}}
\eta(\tau_2)^{1+2N_{D5_1}}\eta(\tau_3)^{1+2N_{D9}}\right]^{-1} e^{2
\pi i [N_{D9}(\epsilon^3+\epsilon^3
\epsilon^6 )+ N_{D5_1} (\epsilon^2+\epsilon^2 \epsilon^5 )+N_{D5_2} (\epsilon^1+\epsilon^1 \epsilon^4 ) ]} \\
\times \prod_{K=1}^{N_{D9}}
\vartheta^2\left[{\frac{1}{2} + \epsilon^6 \atop \frac{1}{2}
    + \epsilon^3 }
\right](\xi^3_K ; \tau_3)  \times \prod_{Q=1}^{N_{D5_1}}
\vartheta^2 \left[{\frac{1}{2} + \epsilon^5 \atop
    \frac{1}{2} +  \epsilon^2 }\right] (\xi^2_Q ; \tau_2)\times \prod_{P=1}^{N_{D5_2}}
\vartheta^2 \left[{\frac{1}{2} + \epsilon^4 \atop
    \frac{1}{2} +  \epsilon^1 }\right] (\xi^1_P ; \tau_1)
\label{deter2}
\end{multline}
where, as in eq.(\ref{3.2.3}), non-holomorphic prefactors have been absorbed in $\hat T_3$ and $\hat \Phi^{\vec i}_{ab}$, according to their definitions, eqs.(\ref{var}) and (\ref{tt}).

Putting (\ref{3.2.3}) and (\ref{deter2}) together, we obtain the full expression for the non-perturbative superpotential in eq.(\ref{i1}),
\begin{multline}
W_{n.p.} = e^{2 \pi i \hat T_3} \hat \Phi_{ab}^{\vec i}\ \
\left[\eta(\tau_1)^{1+2N_{D5_2}} \eta(\tau_2)^{1+2N_{D5_1}}
  \eta(\tau_3)^{1+2N_{D9}} \right]^{-1}\ \times \\ \times
  \prod_{r=1}^3  \sum_{\epsilon^r,\epsilon^{r+3}=
  0,1/2}  \left(e^{2 \pi i [N_{D9}(\epsilon^3+ \epsilon^3
\epsilon^6 )+ N_{D5_1} (\epsilon^2+\epsilon^2
\epsilon^5 )+ N_{D5_2} (\epsilon^1+\epsilon^1
\epsilon^4 ) ]}  e^{- i \pi |I^r_{ab}| \epsilon^r
  \epsilon^{r+3}} \right.  \\
 \vartheta\left[{\frac{i_r}{I^r_{ab}}+  \epsilon^{r+3}
    \atop  I_{ab}^r \epsilon^{r} }\right] (I^r_{bE1}\xi_a^r+I^r_{E1a}\xi^r_b\ ;\
\tau^r|I^r_{ab}|)  \\ \left.
\times \
\prod_{K=1}^{N_{D9}}
\vartheta^2\left[{\frac{1}{2} + \epsilon^6 \atop \frac{1}{2}
    + \epsilon^3 }
\right](\xi^3_K ; \tau_3)  \times \prod_{Q=1}^{N_{D5_1}}
\vartheta^2 \left[{\frac{1}{2} + \epsilon^5 \atop
    \frac{1}{2} +  \epsilon^2 }\right] (\xi^2_Q ; \tau_2)\times \prod_{P=1}^{N_{D5_2}}
\vartheta^2 \left[{\frac{1}{2} + \epsilon^4 \atop
    \frac{1}{2} +  \epsilon^1 }\right] (\xi^1_P ; \tau_1)
\right)   \label{superfinal}
\end{multline}
This expression is holomorphic and periodic under shifts of the D-brane moduli and $U(1)$ gauge transformations, and behaves as a holomorphic modular form of weight $-1$ under $SL(2,\mathbb{Z})$ transformations of the complex structure parameters. Of course, symmetrization with respect to all the orbifold and orientifold operations should be understood. In Section \ref{simple} we will present the application of this expression to a concrete orbifold model.

%
\subsection{Local vs. Global}

From the above analysis we conclude that Euclidean brane instantons may indeed generate non-trivial scalar potentials for the D-brane moduli through the presence of linear superpotential couplings for some charged field. One of the valuable features of this approach is that it leads to global expressions, valid for arbitrarily large VEV's of the D-brane moduli. This allows, therefore, to describe the D-brane dynamics over the full compact space.

Eq.(\ref{superfinal}) is thus given as a sum over Euclidean brane instantons, taking into account their different locations in the compact space and their configuration of discrete Wilson lines. From a local perspective, each term in this sum has two types of zeroes:
\begin{enumerate}
\item[(a)]  Points in the moduli space for which the one-loop determinant vanishes due to the appearance of extra fermionic zero modes which run in the loops.

\item[(b)] Points in the moduli space for which the area of the
  3-point disc scattering amplitude becomes zero, leading to a vanishing Yukawa
  coupling between the two fermionic charged zero modes of the
  instanton and the physical chiral field.
\end{enumerate}

The first of these effects has been extensively discussed in the
literature, starting with \cite{ganor}. From the target space point of
view, extra fermionic zero modes arise from strings stretching between the
instanton and non-magnetized bulk D-branes, which become massless when
the brane sits on top of the instanton. The resulting scalar
potential for the moduli of these non-magnetized branes has important
implications for D-brane inflationary models (see e.g. \cite{berg,maldacena,kachru}).

The second effect, i.e. zeroes of the superpotential which originate in vanishing disc amplitudes, is relatively new and leads to non-trivial scalar potentials also for the moduli of magnetized D-branes. Its physical interpretation was given in the introduction as a force acting on the magnetized D-branes by the singularities where the instantons reside. D-branes move in order to reduce the area of the disc and thus minimize the vacuum energy. In this process the gauge symmetry is generically spontaneously broken, leading to a reduction in the rank of the gauge group.

Whereas these two effects have a very different origin,
the local driving force which results from them is the same in both cases:
in models with linear terms, D-branes are subject to some attractive or repulsive force from
Euclidean brane instantons.

This simple picture is, however, further complicated by global issues.
As we have just commented, in a compact model there are usually several locations on the
internal space where instantons reside. Each of these loci exerts a force on
the D-branes. The latter therefore tend to get stabilized at points where all non-perturbative forces are balanced. Thus, supersymmetric vacua in global models are generically the result of a combined cancelation between different non-null instanton contributions.

We can illustrate more explicitly this phenomenon with a simple toy model. Consider a toroidal compactification with D$5_3$-branes equally distributed and sitting on top of the E1-instantons at the singularities. In this case E1-instantons correspond to standard QFT instantons for the gauge theory in the worldvolume of the D$5_3$-branes. In addition, consider that there are also four non-magnetized bulk D$5_1$-branes in the model, i.e. $N_{D5_1}=4$ and $N_{D9}=N_{D5_2}=0$ in eq.(\ref{deter2}). The induced non-perturbative superpotential is in this case \cite{ccd},
\begin{equation}
W_{n.p.} =
e^{2 \pi i \hat T_3}\left[\eta(\tau_1)^{1}
\eta(\tau_2)^{9}\eta(\tau_3)^{1}\right]^{-1} \sum_{\epsilon^2,\epsilon^{5}=
  0,1/2}\prod_{q=1}^{4}
\vartheta^2 \left[{\frac{1}{2} + \epsilon^5 \atop
    \frac{1}{2} +  \epsilon^2 }\right] (\xi^2_q ; \tau_2)\label{toyglobal}
\end{equation}
where $\xi^2_q$ are the positions of the D$5_1$-branes on the second torus.

Let us analyze the non-perturbative dynamics for the D$5_1$-branes. From previous local considerations (point (a) above) it is clear that when the D$5_1$-branes sit on top of the instantons, i.e.
\begin{equation}
\xi^2_1=0 \ , \qquad \xi^2_2=\frac12 \ , \qquad \xi^2_3=\frac{\tau_2}{2}\ , \qquad \xi^2_4=\frac12+\frac{\tau_2}{2}\ ,\label{susy0}
\end{equation}
each term in the sum over instantons in eq.(\ref{toyglobal}) vanishes. We therefore expect a non-perturbative supersymmetric vacuum ($\langle F_{\hat T_3}\rangle=0$) at this point of the moduli space. The gauge symmetry is $U(1)^8$. Outside this locus one or more E1-instantons have the right number of neutral zero modes to give a non-null contribution to the superpotential. Hence, from this local analysis we would conclude that (\ref{susy0}) is a supersymmetric minimum of the scalar potential, with no flat directions.

\begin{figure}[!h]
\begin{center}
\includegraphics[width=6cm]{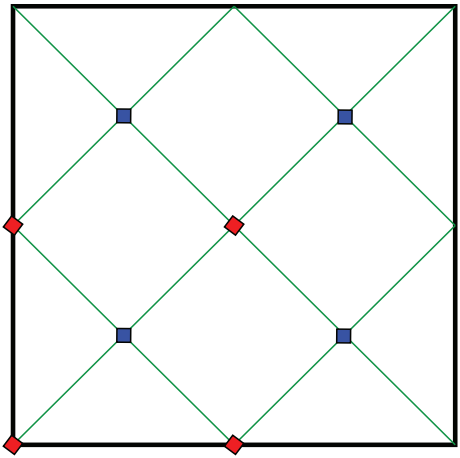}
\end{center}
\caption{\label{figtoy} Non-perturbative supersymmetric flat
  directions for the motion of D$5_1$-branes along the second
torus. The branes can only move along the continuous lines, where
gauge group is $USp(2)^4$. Points with special gauge
symmetry group have been also marked: (blue) squares for $USp(4)^2$ points, and (red) rhombus for $U(1)^8$ points. The four branes can only move in a correlated way, so that they approach an $USp(4)^2$ point all at the same time, each one through a different direction.}
\end{figure}

The global analysis, however, reveals that the vacuum structure is actually more complicated. Even if each single instanton contribution in eq.(\ref{toyglobal}) is non-zero, there can be cancelations between different terms leading to new supersymmetric vacua. In the toy model at hand one may check that eq.(\ref{susy0}) actually belongs to a one-parameter family of supersymmetric vacua,
\begin{equation}
\xi^2_1=\frac{\rho}{2}+\frac{\rho\tau_2}{2} \ , \quad \xi^2_2=\frac{1-\rho}{2}+\frac{\rho\tau_2}{2} \ , \quad \xi^2_3=\frac{\rho}{2}+\frac{(1-\rho)\tau_2}{2}\ , \quad \xi^2_4=\frac{1-\rho}{2}+\frac{(1-\rho)\tau_2}{2}\label{susy1}
\end{equation}
with $\rho\in[0,1)$. Plugging (\ref{susy1}) into (\ref{toyglobal}) leads to $W_{n.p.}=\langle F_{\hat T_3}\rangle=0$.

We have represented in Figure \ref{figtoy} the supersymmetric locus along which D$5_1$-branes can move without any cost of energy. Interestingly, $D5_1$-branes move in a correlated way, accordingly to a $\mathbb{Z}_4$ discrete symmetry which has emerged non-perturbatively. At the special point $\rho=0$ the theory is in the $U(1)^8$ phase and the supersymmetric vacuum can be explained in terms of extra neutral instanton zero modes, as before. Outside this locus, the vacuum can only be understood in terms of interactions between different instanton effects. The gauge group on the D$5_1$-branes becomes $USp(2)^4$, except at the locus $\rho=1/2$ on which the gauge symmetry is enhanced to $USp(4)^2$.

This simple model thus reveals that the
global non-perturbative D-brane dynamics can in general differ substantially from the local analysis. Similar statements
can be equally made for stringy instanton effects (point (b)
above). In this case, however, there are some differences. Indeed, we have
noted that there are $|I_{ab}|$ F-terms, corresponding to different degenerate charged fields. Whereas, for each of these
degenerate fields, it is in principle possible to set $\langle F_\Phi\rangle=0$ by changing the VEV's of the
Wilson line moduli of branes $a$ and $b$, from eq.(\ref{3.2.3}) we see
that the value for which this occurs is different for each degenerate
field (it depends on the value of $i_r$ in eq.(\ref{3.2.3})). Thus, in general, in toroidal models it is not possible to set to zero
all the F-terms by this mechanism.

\begin{figure}[!h]
\begin{center}
\includegraphics[width=13cm]{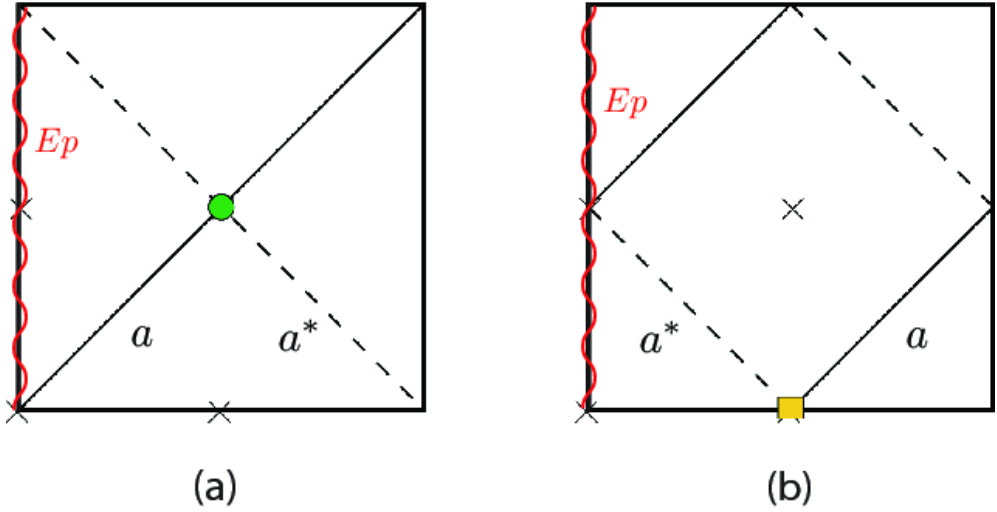}
\end{center}
\caption{\label{otra} Brane positions for which the F-term of one of the two degenerate antisymmetric fields in a $(m,n)=(1,1)$ toy model vanishes. We have represented with a waving (red) line the instanton. Antisymmetrics are localized at the intersection between brane $a$ and its orientifold image $a^*$, and are denoted by a square and a circle.}
\end{figure}

The situation is perhaps better
depicted in the type IIA T-dual with intersecting D6-branes. Consider
for instance a model where Euclidean brane instantons
induce linear superpotential couplings for fields in the antisymmetric
representation of the $U(2)$ gauge symmetry of a brane. For
simplicity, let us take this brane to have magnetization $(m,n)=(1,1)$
along one of the 2-torus. We have represented in Figure \ref{otra} the
corresponding configuration in terms of D6-branes. There are two degenerate antisymmetrics,
localized at each of the two intersections between the brane and its orientifold/orbifold image. The position of the
brane for which the relevant Yukawa coupling in the instanton action
vanishes has been
represented in Figures \ref{otra} (a) and (b). These positions are indeed incompatible with each other, so a supersymmetric minimum cannot be achieved by stabilizing the position of the brane. Nevertheless, combining the
contribution of both types of F-terms to the scalar potential, we
observe that there is a minimum at the symmetric position with
respect to all the singularities, where all forces are balanced.

\section{Global model with brane recombination}
\label{simple}

In this Section we apply the above ideas in a fully consistent model.
For that, we consider compactifications of
type I string theory on $T^6/\mathbb{Z}_2\times \mathbb{Z}_2$ with discrete torsion \cite{cristinatorsion,blumenrigid,mikerigid}, where
the $\mathbb{Z}_2\times \mathbb{Z}_2$ acts in the way described in
Section \ref{stringy}.

As we have already commented, one can distinguish various types of branes in these models. Bulk branes are neutral under some of the twisted fields and therefore can move in the bulk. Branes which are charged under a single twisted sector can only move on a $T^2$, whereas branes with no twisted charge of any kind can move in the full compact space. On the other hand, fractional branes carry a non-trivial charge of all kind of twisted fields and therefore are stuck at the singularities.

The twisted charge of a given stack of fractional branes can be parameterized by the $64$ choices of discrete Wilson lines, described in Section \ref{stringy}, and four inequivalent choices for the Chan-Paton charge. Here, following \cite{mikerigid}, we consider two particular choices of Chan-Paton charges. These are given in terms of complex charges $p_a$ and $q_\alpha$ as,
\begin{equation}
\begin{split}
&N_{a,o}=p_a+\bar p_a \,,
\\
&N_{a,g}=i(p_a-\bar p_a) \,,
\\
&N_{a,f}=i(p_a-\bar p_a)\,,
\\
&N_{a,h}=p_a+\bar p_a\,,
\\
\end{split}\qquad
\begin{split}
&N_{\alpha,o}=q_{\alpha}+\bar q_{\alpha} \,,
\\
&N_{\alpha,g}=i(q_{\alpha}-\bar q_{\alpha})\,,
\\
&N_{\alpha,f}=-i(q_{\alpha}-\bar q_{\alpha})\,,
\\
&N_{\alpha,h}=-q_{\alpha}-\bar q_{\alpha} \,,
\\
\end{split}
\label{cp}
\end{equation}
The resulting gauge group is the product of unitary factors
\begin{equation}
G_{\rm CP} = \prod_aU(p_a)\times\prod_{\alpha} U(q_{\alpha}) \,.
\end{equation}
The remaining types of fractional branes can be obtained by considering branes with the above choices of Chan-Paton charge, but magnetization $(m^r,n^r)\to(-m^r,n^r)$, $r=1,2,3$.

We summarize the spectrum of charged chiral fields in Table \ref{table1}, where the orientifold acts in the usual way on the magnetization numbers,
\begin{align*}
a :  & (m^1,n^1)\otimes (m^2,n^2)\otimes (m^3,n^3)\\
a^* :  & (-m^1,n^1)\otimes (-m^2,n^2)\otimes (-m^3,n^3)\nonumber
\end{align*}
The intersection between the brane and the orientifold planes is defined as,
\begin{equation}
I_{AO}=8\left(m^1_{A}m^2_{A}m^3_{A}- m^1_{A}n^2_{A}n^3_{A}- n^1_{A}m^2_{A}n^3_{A}
+ n^1_{A}n^2_{A}m^3_{A}\right)
\end{equation}

The whole set of D-branes and magnetization numbers in a
given model have to satisfy stringent conditions
coming from supersymmetry and cancelation of tadpoles and
non-homological K-theory charges. These conditions can be
found e.g. in \cite{blumenrigid,mikerigid}. For completeness we also summarize them in Appendix \ref{KTheory}.

\begin{table}[!htb]
\begin{center}
\begin{tabular}{|c|c|c|}
\hline
Multiplicity & Representation & Relevant Indices \\
\hline
\hline
$\frac{1}{8}(I_{aa^*}+I_{aO}-4I^1_{aa^*}-4I^2_{aa^*}+4I^3_{aa^*})$ & $\left(\frac{p_a(p_a-1)}{2},1\right)$ & $\forall a$
\\[1ex]
$\frac{1}{8}(I_{\alpha\alpha^*}+I_{\alpha O}-4I^1_{\alpha\alpha^*}-4I^2_{\alpha\alpha^*}+4I^3_{\alpha\alpha^*})$ &
$\left(1,\frac{q_{\alpha}(q_{\alpha}-1)}{2}\right) $&$\forall \alpha$
\\[1ex]
$\frac{1}{8}(I_{aa^*}-I_{aO}-4I^1_{aa^*}-4I^2_{aa^*}+4I^3_{aa^*})$ & $\left(\frac{p_a(p_a+1)}{2},1\right)$ & $\forall a$
\\[1ex]
$\frac{1}{8}(I_{\alpha\alpha^*}-I_{\alpha O}-4I^1_{\alpha\alpha^*}-4I^2_{\alpha\alpha^*}+4I^3_{\alpha\alpha^*})$ &
$\left(1,\frac{q_{\alpha}(q_{\alpha}+1)}{2}\right) $&$\forall \alpha$
\\[1ex]
\hline
$\frac{1}{4}(I_{a\alpha^*}-S_g^{a\alpha}I^1_{a\alpha^*}+S_f^{a\alpha}I^2_{a\alpha^*}-S_h^{a\alpha}I^3_{a\alpha^*})$& $(p_a,q_{\alpha})$ & $\forall a,\forall\alpha$
\\[1ex]
$\frac{1}{4}(I_{a\alpha}+S_g^{a\alpha}I^1_{a\alpha}-S_f^{a\alpha}I^2_{a\alpha}-
S_h^{a\alpha}I^3_{a\alpha})$& $(p_a,\bar{q}_{\alpha})$ &$\forall a,\forall \alpha$
\\[1ex]
\hline
$\frac{1}{4}(I_{ab^*}-S_g^{ab}I^1_{ab^*}-S_f^{ab}I^2_{ab^*}+S_h^{ab}I^3_{ab^*})$ & $(p_a,p_b)$ & $a < b$
\\[1ex]
$\frac{1}{4}(I_{ab}+S_g^{ab}I^1_{ab}+S_f^{ab}I^2_{ab}+S_h^{ab}I^3_{ab})$& $(p_a,\bar{p}_b)$& $a < b$
\\[1ex]
$\frac{1}{4}(I_{\alpha\beta^*}-S_g^{\alpha\beta}I^1_{\alpha\beta^*}-S_f^{\alpha\beta}I^2_{\alpha\beta^*}
+S_h^{\alpha\beta}I^3_{\alpha\beta^*})$ & $(q_{\alpha},q_{\beta})$& $\alpha < \beta$
\\[1ex]
$\frac{1}{4}(I_{\alpha\beta}+S_g^{\alpha\beta}I^1_{\alpha\beta}+S_f^{\alpha\beta}I^2_{\alpha\beta}
+S_h^{\alpha\beta}I^3_{\alpha\beta})$&  $(q_{\alpha},\bar{q}_{\beta})$ & $\alpha < \beta $
\\[1ex]
\hline
1 & $(p_a,\bar{q}_\alpha) + (\bar{p}_a, q_\alpha)$ & if $(m^i_a,n^i_a) = (m^i_\alpha,n^i_\alpha)~\forall i$
\\[1ex]
\hline
1 & $(p_a, q_\alpha) + (\bar{p}_a, {\bar q}_\alpha)$ & if $(m^i_a,n^i_a) = (-m^i_\alpha,n^i_\alpha)~\forall i$
\\[1ex]
\hline
\end{tabular}
\end{center}
\caption{Representations and multiplicities of charged chiral
superfields on a $T^6 /\mathbb{Z}_2 \times \mathbb{Z}_2$ orbifold
with discrete torsion, in the presence of magnetic backgrounds. $S^{AB}_{i=g,f,h}$ denotes the number of fixed points of the
generator $i=g,f,h$ that both branes A \ and B intersect.}
\label{table1}
\end{table}

\subsection{The model}

The particular model considered in this Section was first presented in Ref.~\cite{mikerigid}. It consists of
four stacks of fractional D9-branes with identical magnetization,
\begin{equation}
\begin{split}
1\ :\qquad &(m^i,n^i)= (-2,1)\otimes (-1,1) \otimes (1,1) \,,
\\
2\ :\qquad &(m^i,n^i)= (-2,1)\otimes (-1,1) \otimes (1,1) \,,
\\
3\ : \qquad&(m^i,n^i)= (2,1) \otimes (1,1) \otimes (-1,1) \,,
\\
4\ : \qquad&(m^i,n^i)= (2,1) \otimes (1,1) \otimes (-1,1) \,,
\end{split}\label{frac}
\end{equation}
Branes $1$ and $3$ are of type $(p_a,q_\alpha)=(2,0)$ in eq.(\ref{cp}), whereas branes $2$ and $4$ are of type $(p_a,q_\alpha)=(0,2)$. This setup has vanishing total twisted and non-homological K-theory charges. In particular, recombining the four fractional branes leads to a  magnetized bulk brane, as it will be described in the next subsection.

Cancelation of untwisted tadpoles require, in addition,
four non-magnetized D9-branes
and four D$5_1$-branes to be present in the model, of wrapping numbers
\begin{equation}
\begin{split}
c\ :\qquad &(m_i^c,n_i^c)= (0,1) \otimes (0,1) \otimes (0,1)  \,,
\\
d\ : \qquad &(m_i^d,n_i^d)= (0,1) \otimes (1,0) \otimes (-1,0)  \,.
\\
\end{split}
\end{equation}
In order to simplify the formalism as much as possible, we have considered $D5$-branes as particular cases of magnetized D9-branes. This interpretation is particularly natural in the light of the type IIA dual picture, where both objects become D6-branes.

The gauge group which results is,
\begin{equation}
G = U(2)^2\times U(2)^2 \times USp(4)^2 \times USp(4)^2\,
\label{cpgroup}
\end{equation}
Each pair of unitary groups live on a stack of fractional branes,
while the symplectic factors originate from the non-magnetized D9 and
D5-branes that, as anticipated, can be displaced in the bulk. Charged matter corresponds to chiral supermultiplets transforming in bi-fundamental and antisymmetric representations of the gauge group. We have listed in Table \ref{spectrum3} the part of the spectrum which is charged under the unitary factors of the gauge group.
Aside from these modes, the model also includes open strings stretched between non-magnetized (D9 and/or D5) branes and open strings stretched between non-magnetized branes and fractional (magnetized) ones. Their generic massless excitations are omitted for space considerations but can be found in \cite{mikerigid}.

\begin{table}[htb!]
\begin{center}
\begin{tabular}{|c|c|c|}
\hline
Multiplicity & Representation & field
\\[.5ex]
\hline
\hline
12 & $(\bar1,1,1,1)$ & $A^1$
\\[.3ex]
12 & $(1,\bar1,1,1)$ & $A^2$
\\[.3ex]
12& $(1,1,{\bar 1},1)$ & $A^3$
\\[.3ex]
12& $(1,1,1,{\bar 1})$ & $A^4$
\\[.3ex]
\hline
4&$(\bar2,\bar2,1,1)$ & $\Phi_{\bar1\bar2}$
\\[.3ex]
4 & $(1,1,2,2)$ & $\Phi_{34}$
\\[.3ex]
\hline
1 & $(2,\bar2,1,1)+(\bar2,2,1,1)$ & $\Phi_{1\bar2}+\tilde\Phi_{\bar12}$
\\[.3ex]
1&$(1,1,2,\bar2)+(1,1,\bar2,2)$& $\Phi_{3\bar4}+\tilde\Phi_{\bar34}$
\\[.3ex]
1&$(2,1,1,2)+(\bar 2,1,1,\bar 2)$& $\Phi_{14}+\tilde\Phi_{\bar1\bar4}$
\\[.3ex]
1&$(1,2,2,1)+(1,\bar2,\bar2,1)$& $\Phi_{23}+\tilde\Phi_{\bar2\bar3}$
\\[.3ex]
1&$(2,1,2,1)+(\bar2,1,\bar2,1)$& $\Phi_{13}+\tilde\Phi_{\bar1\bar3}$
\\[.3ex]
1&$(1,2,1,2)+(1,\bar2,1,\bar2)$& $\Phi_{24}+\tilde\Phi_{\bar2\bar4}$
\\[.3ex]
\hline
\end{tabular}
\end{center}
\caption{Charged massless spectrum with respect to the $U(2)^2 \times U(2)^2$ subgroup of the $U(2)^2 \times U(2)^2 \times USp(4)^2 \times USp(4)^2$ gauge group, where the field $A^i$ is in the antisymmetric representation of the $i$-th $SU(2)$ factor. To lighten the notation we have not explicitly listed the $U(1)$ charges. They can be easily derived from the $U(2)$ representations.}
\label{spectrum3}
\end{table}


\subsection{D-brane recombination and Higgsing}
\label{higgs}

Let us describe in more detail the recombination of the four stacks of fractional branes (\ref{frac}). More generically, consider four stacks of fractional branes at an orbifold singularity with,
\begin{equation}
\begin{split}
1\ :\qquad &(m^1,n^1)\otimes (m^2,n^2) \otimes (m^3,n ^3) \,,
\\
2\ :\qquad &(m^1,n^1)\otimes (m^2,n^2) \otimes (m^3,n ^3) \,,
\\
3\ : \qquad&(-m^1,n^1)\otimes (-m^2,n^2) \otimes (-m^3,n ^3) \,,
\\
4\ : \qquad&(-m^1,n^1)\otimes (-m^2,n^2) \otimes (-m^3,n ^3) \, .
\end{split}
\label{r1}
\end{equation}
Branes $1$ and $3$ are of type $(p_a,q_\alpha)=(s,0)$ in eq.(\ref{cp}), whereas branes $2$ and $4$ are of type $(p_a,q_\alpha)=(0,s)$. The gauge group is $U(s)_1\times U(s)_2\times U(s)_3\times U(s)_4$. Notice in particular that the twisted and untwisted charges of the system are exactly the same as that of a bulk $U(s)$ brane, sitting in an arbitrary position in the internal space. It is actually straightforward to prove that this system of fractional branes can recombine into the corresponding bulk brane,
by giving VEV's to fields transforming in the bifundamental representation of the gauge group. These VEV's correspond to the positions of the bulk brane after recombination. To check this,
we start from the spectrum in the $U(s)^4$ gauge theory.
It contains all type of fields transforming in bifundamental,
antisymmetric and symmetric representations under all gauge
factors. Among the bifundamentals, there are twelve complex fields,
\begin{align}
\Phi_{1\bar2}\ : & \ (s,\bar s,1,1) & \tilde\Phi_{\bar12}\ : & \ (\bar s,s,1,1) &
\Phi_{3\bar4}\ : & \ (1,1,s,\bar s) & \tilde\Phi_{\bar34}\ : & \ (1,1,\bar s,s) \\
\Phi_{14}\ : & \ (s,1,1,s) & \tilde\Phi_{\bar1\bar4}\ : & \ (\bar s,1,1,\bar s) &
\Phi_{23}\ : & \ (1,s,s,1) & \tilde\Phi_{\bar2\bar3}\ : & \ (1,\bar s,\bar s,1) \\
\Phi_{13}\ : & \ (s,1,s,1) & \tilde\Phi_{\bar1\bar3}\ : & \ (\bar s,1,\bar s,1) &
\Phi_{24}\ : & \ (1,s,1,s) & \tilde\Phi_{\bar2\bar4}\ : & \ (1,\bar s,1,\bar s)
 \label{r2}
\end{align}
After turning on VEV's for these fields, the complex Wilson line moduli $\xi^i_a$, $i=1,2,3$, of the recombined bulk brane are identified as
\begin{eqnarray}
&& \Phi_{1\bar2}  = \tilde\Phi_{\bar12} = \xi^1_{\alpha,x} \ , \qquad \Phi_{3\bar4} =
\tilde\Phi_{\bar34} = \xi^1_{\alpha,y} \ , \nonumber \\
&& \Phi_{13} = \tilde\Phi_{\bar1\bar3} = \xi^2_{\alpha,x} \ , \qquad \Phi_{24} =
\tilde\Phi_{\bar2\bar4} = \xi^2_{\alpha,y} \ , \nonumber \\
&& \Phi_{14} = \tilde\Phi_{\bar1\bar4} = \xi^3_{\alpha,x} \ , \qquad \Phi_{23} =
\tilde\Phi_{\bar2\bar3} = \xi^3_{\alpha,y} \ , \nonumber \\
&&\qquad \quad \hat \xi^r_{\alpha} \ = \ \xi^r_{\alpha,x} \ + \ \tau_r \ \xi^{r}_{\alpha,y}  \label{r3}
\end{eqnarray}
Hence Wilson lines correspond to the three chiral fields which are
left massless in the Higgsing $U(s)^4 \to U(s)$, whereas the
remaining nine fields are eaten up, giving masses to the broken $U(s)^3$ generators.

Apart from the above fields transforming in bifundamental representations, there are also fields transforming in
antisymmetric and symmetric representations of the gauge group with multiplicities (see also Table \ref{table1}),
\begin{align}
N_A^{\rm fr} & =  \frac{1}{8}(I_{aa^*} + I_{aO} - 4 I^1_{aa^*} - 4 I^2_{aa^*} + 4 I^3_{aa^*}) \ , \label{nafr}\\
N_S^{\rm fr} & =  \frac{1}{8}(I_{aa^*} - I_{aO} - 4 I^1_{aa^*} - 4 I^2_{aa^*} + 4 I^3_{aa^*}) \ ,\label{nsfr}
\end{align}
respectively. After recombination, these become part of the fields transforming in antisymmetric or symmetric representations of the bulk brane, with multiplicities
\begin{equation}
N_A \ = \ 4 N_{A}^{\rm fr} + N_{A}^{\rm bi} \ =\ 2 I_{aa^*} \ + \ \frac{1}{2} I_{aO} \quad , \quad N_S \ = \ 4 N_{S}^{\rm fr} + N_{S}^{\rm bi} \ =\ 2 I_{aa^*} \ - \ \frac{1}{2} I_{aO} \ . \label{r4}
\end{equation}
where $N_{A,S}^{\rm bi}$ is the contribution to the multiplicities for the bulk
(anti)-symmetrics arising from charged
bifundamentals via brane recombination. More precisely,
\begin{equation}
N_{A}^{\rm bi} \ =  \ N_{S}^{\rm bi} \ = \ \frac{3}{2} I_{aa^*} +  2 ( I^1_{aa^*} + I^2_{aa^*} - I^3_{aa^*}) \ . \label{r5}
\end{equation}
We therefore succeeded to follow precisely the recombination of the
fractional branes (\ref{r1}) into a bulk brane from the point of view of Higgs mechanism in the effective gauge theory.

Hence, in the particular model eq.(\ref{frac}), by giving VEV's to the bilinears $\langle \Phi_M\tilde\Phi_{\bar M}\rangle$, fractional D9-branes $1,\ldots,4$ can recombine into a bulk brane $a$ and move along one or more $T^2$ in the $T^6/\mathbb{Z}_2\times \mathbb{Z}_2$. In the most general case where all the bilinears $\langle \Phi_M\tilde\Phi_{\bar M}\rangle$ get a VEV, the gauge group in eq.(\ref{cpgroup}) is broken to,
\begin{equation}
U(2)^2\times U(2)^2\times USp(4)^2 \times USp(4)^2\to U(2)_{\rm diag.}\times USp(4)^2 \times USp(4)^2\label{u2break}
\end{equation}
After recombination, the fields $\langle \Phi_M\tilde\Phi_{\bar
  M}\rangle$ become the degrees of freedom of three complex
scalars transforming in the adjoint representation of $U(2)_{\rm
  diag.}$, according to eq.(\ref{r3}), and parameterizing the continuous Wilson line
deformations of the recombined brane. Each of the chiral fields $\Phi_{\bar1\bar2}$
and $\Phi_{34}$ leads to fields in the symmetric and the antisymmetric
representation of $U(2)_{\rm diag.}$. Thus, the
chiral spectrum of the recombined brane $a$ consists of $4\times 14=56$ antisymmetrics and $4\times 2=8$ symmetrics, which we denote, respectively, as $A^{M,(i_2,i_3)}$, $M=1\ldots 14$, and $S^{Q,(i_2,i_3)}$, $Q=1,2$, with $i_{2,3}=0,1$.

In what follows we will see that instanton effects in this model can actually trigger this recombination through the generation of linear superpotential couplings for the fields $A^{M,(i_2,i_3)}$.


\subsection{Non-perturbative D-brane dynamics}

We can now turn to the analysis of the superpotential induced by rigid ${\rm E}1$-brane instantons. The twisted charge of the instanton is parameterized by the 64 possible choices for the discrete Wilson lines and positions described in Section \ref{stringy}, plus four different choices for the Chan-Paton charge. In this case we choose to parameterize the latter as,
\begin{align}
D_{1,o}&=k_1 & D_{2,o}&=k_2 & D_{3,o}&=k_3 & D_{4,o}&=k_4 \nonumber\\
D_{1,g}&=k_1 & D_{2,g}&=k_2 & D_{3,g}&=-k_3 & D_{4,g}&=-k_4 \nonumber\\
D_{1,f}&=-k_1 & D_{2,f}&=k_2 & D_{3,f}&=k_3 & D_{4,f}&=-k_4 \nonumber\\
D_{1,h}&=-k_1 & D_{2,h}&=k_2 & D_{3,h}&=-k_3 & D_{4,h}&=k_4\label{E1cp}
\end{align}
For $O(1)$ instantons $k_i=1$ (see \cite{mikerigid} for more details). If non-magnetized branes are a distance away from the singularities, ${\rm E}1$-branes localized at the orbifold singularities have the right structure to generate linear couplings in the superpotential for the antisymmetrics. We list in Table \ref{finalinst} the spectrum of charged fermionic zero modes.

\begin{table}[htb]
\begin{center}
\begin{tabular}{|c|c|c|}
\hline
CP choice & Representation & Zero mode
\\[1ex]
\hline
\hline
$1$ & $(2,1,1,1,1,1,1,1)$ & $\eta^{1}_i$
\\[1ex]
$2$ & $(1,2,1,1,1,1,1,1)$ & $\eta^{2}_i$
\\[1ex]
$3$ & $(1,1,\bar 2,1,1,1,1,1)$ & $\eta^{3}_i$
\\[1ex]
$4$ & $(1,1,1,\bar 2,1,1,1,1)$ & $\eta^{4}_i$
\\[1ex]
\hline
\end{tabular}
\end{center}
\caption{Charged zero-mode structure for $O(1)$ E1-brane instantons in the $U(2)^2\times U(2)^2\times USp(4)^2 \times USp(4)^2$ model. The index $i$ refers to the (anti)fundamental representation of $U(2)$.}
\label{finalinst}
\end{table}

Hence, consider the gauge invariant action of a single instanton with Chan-Paton charge of the first type in Table \ref{finalinst}. Before recombination of branes (\ref{frac}), the action has the expression,
\begin{equation}
S_{\rm inst} = S_{{\rm E}1} + \sum_{i,j=1}^2\left[f_1[\Phi\tilde\Phi,\xi]\ \eta^{1}_i \, A^{1}_{ij}\, \eta^{1}_j \ + \ g_1[\Phi\tilde\Phi,\xi]\ \sum_{k=1}^2\eta^{1}_i \, (\tilde\Phi_{\bar 1 2})_{ki}\, (\Phi_{\bar1\bar2})_{kj}\, \eta^{1}_j \right]
\end{equation}
where $f_1[\Phi\tilde\Phi,\xi]$ and $g_1[\Phi\tilde\Phi,\xi]$ are functions of the bilinears $\Phi_M\tilde\Phi_{\bar M}$ in Table \ref{spectrum3} and of the adjoint scalars which parameterize the position and Wilson line deformations of the non-magnetized bulk D9 and D5-branes. For simplicity, we have omitted indices running over the multiplicities of the fields. Similar actions can be written for the three other types of Chan-Paton choices in Table \ref{finalinst}.

Summing over all possible instanton configurations and integrating over the instanton charged zero modes $\eta^{M}_i$, $M=1,\ldots,4$, the following non-perturbative contribution to the superpotential arises
\begin{multline}
W_{\rm n.p.} \, = \,  \sum_{\alpha}\sum_{i,j=1}^2\epsilon_{ij} \,\left[\sum_{M=1}^4e^{- S_{{\rm E}1_{\alpha,M}}} \,  f_M[\Phi\tilde\Phi,\xi]\ A_{ij}^M\ \right.\\
+\ \sum_{k=1}^2 \left(e^{- S_{{\rm E}1_{\alpha,1}}}g_1[\Phi\tilde\Phi,\xi](\tilde\Phi_{\bar 1 2})_{ki}+e^{- S_{{\rm E}1_{\alpha,2}}}g_2[\Phi\tilde\Phi,\xi](\Phi_{1\bar2})_{ki}\right)\, (\Phi_{\bar1\bar2})_{kj}\ \\ \left. +\ \sum_{k=1}^2 \left(e^{- S_{{\rm E}1_{\alpha,3}}}g_3[\Phi\tilde\Phi,\xi](\Phi_{3 \bar 4})_{ki}+e^{- S_{{\rm E}1_{\alpha,4}}}g_4[\Phi\tilde\Phi,\xi](\tilde\Phi_{\bar3 4})_{ki}\right)\, (\Phi_{34})_{kj}\right]
\label{sys6}
\end{multline}
where $g_2$, $g_3$ and $g_4$ are generated by the other type of instantons in Table \ref{finalinst}.
Bilinears $\langle\Phi_M\tilde\Phi_{\bar M}\rangle$ and the
moduli of the non-magnetized bulk D9 and D5-branes may
acquire a VEV in order to minimize the vacuum energy.

For arbitrary VEV's $\langle\Phi_M\tilde\Phi_{\bar M}\rangle$ the four stacks of fractional branes, eq.(\ref{frac}), recombine, as described in Section \ref{higgs}. The linear (in the charged fields) superpotential after complete recombination is given by eq.(\ref{superfinal}) with $N_{D9}=N_{D5_1}=4$. After symmetrization with respect to all the orbifold and orientifold generators, the resulting superpotential reads,
\begin{equation}
W_{\rm n.p} \ = \ \left[\eta(\tau^1) \eta(\tau^2)^{3}
  \eta(\tau^3)^{3} \right]^{-1} \  \sum_{i_2,i_3=0,1}\sum_{M=1}^{14}
e^{2\pi i \hat T_3}\hat{A}^{M,(i_2,i_3)}\ F_{A^{M,(i_2,i_3)}}\label{otrosuper}
\end{equation}
Notice in particular that the 8 fields transforming in symmetric
representations after recombination do not get a linear term. The same conclusion could have
been reached by gauge invariance of the instanton action. On the contrary, the 56 antisymmetrics in the massless spectrum after recombination appear linearly in the superpotential. The explicit expression for the functions $F_{A^{M,(i_2,i_3)}}$ can be found in Appendix \ref{sym}. These factorize in products of three functions, describing the motion of the branes on each of the 2-tori.

 The superpotential (\ref{otrosuper}) preserves an R-symmetry. Note, in particular, that the number of complex fields on which the functions $F_{A^{M,(i_2,i_3)}}$ depend (36, neglecting the complex structure moduli) is much lower than the multiplicity of the antisymmetrics. Hence, according to the general arguments in \cite{seiberg1,seiberg2}, if $F_{A^{M,(i_2,i_3)}}$ were arbitrary functions the above superpotential would spontaneously break the supersymmetry. We may show, however, that this is not the case. The reason is that there are strong correlations between the zeroes of the functions $F_{A^{M,(i_2,i_3)}}$.

Let us be more precise. We can easily see from the expressions in Appendix \ref{sym} that the F-terms associated to 12 antisymmetrics ($\langle F_{A^{4,(i_2,i_3)}}\rangle$, $\langle F_{A^{8,(i_2,i_3)}}\rangle$ and $\langle F_{A^{14,(i_2,i_3)}}\rangle$) are independent of the moduli of the recombined brane, and therefore of the VEV of the bilinears $\Phi_M\tilde\Phi_{\bar M}$. In the T-dual compactification with intersecting D6-branes this can be understood geometrically by noticing that some of the antisymmetrics cannot be moved out of the worldvolume of
the instanton by displacing the recombined brane. A good strategy for looking at supersymmetric minima
($\partial_{A^{M,(i_2,i_3)}}W_{\rm  n.p}=W_{\rm n.p}=0$) is then to
first looking for zeroes of these 12 F-terms, fixing in this way the
moduli of the non-magnetized bulk D9-branes and/or
D5-branes. From eqs.(\ref{ff}) and (\ref{psi4}) in the Appendix, one
may check that the zeroes correspond to one of the following situations:
\begin{itemize}
\item The discrete Wilson lines of non-magnetized bulk D9-branes in the third 2-torus are equally distributed among the four possible choices
    \begin{equation}
\xi^3_1=0 \ , \quad \xi^3_2=\frac{1}{2} \ , \quad \xi^3_3=\frac{\tau_3}{2}\ , \quad \xi^3_4=\frac{1}{2}+\frac{\tau_3}{2}\label{vacua1}
\end{equation}

\item Bulk D5-branes are equally distributed among the four singularities in the second 2-torus
    \begin{equation}
\phi^2_1=0 \ , \quad \phi^2_2=\frac{1}{2} \ , \quad \phi^2_3=\frac{\tau_2}{2}\ , \quad \phi^2_4=\frac{1}{2}+\frac{\tau_2}{2}\label{vacua2}
\end{equation}
\end{itemize}
In any of these two cases extra instantonic zero modes are generated and one may check that all the remaining 44 F-terms also vanish, leading to a non-perturbative supersymmetric vacuum.\footnote{One might worry about potential run-away directions for the closed string moduli. It is possible to show, however, that if eqs.(\ref{vacua1}) or (\ref{vacua2}) are satisfied, the F-terms of the complex structure and K\"ahler moduli also vanish, with $\langle W\rangle=0$. The VEV of these moduli remain as flat directions of the low energy effective action. In addition, D-term conditions are satisfied in the usual way.}

The VEVs of the Wilson line moduli of magnetized D9-branes in this model remain as flat directions of the scalar potential in the non-perturbative supersymmetric vacuum. From the point of view of the gauge theory, the gauge symmetry associated to non-magnetized bulk D9-branes or D5-branes is necessarily spontaneously broken at the non-perturbative vacuum,
\begin{equation}
U(2)^2\times U(2)^2 \times USp(4)^2 \times USp(4)^2\ \to \ U(2)^2\times U(2)^2 \times U(1)^8 \times USp(4)^2
\end{equation}

\begin{figure}[!h]
\begin{center}
\includegraphics[width=13cm]{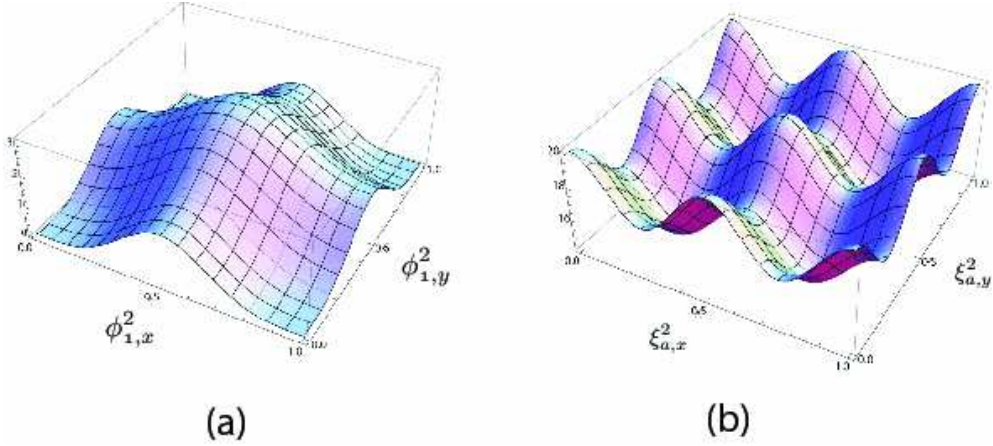}
\end{center}
\caption{\label{junk-potential} (a) Non-perturbative scalar potential for the complex field $\phi^2_1$, parameterizing the position of one of the D5-branes along the second 2-torus. Other D5-branes are equally distributed over the singularities in the second 2-torus, except for the origin. (b) Non-perturbative scalar potential for the complex Wilson line modulus of the recombined D9-brane along the second 2-torus, for an arbitrary VEV of $\phi^2_1$.}
\end{figure}

Fluctuations of non-magnetized bulk D9-branes or D5-branes moduli
around this vacuum, however, induce a non-trivial
scalar potential also for the Wilson line moduli of magnetized
D9-branes. We have plotted in Figure \ref{junk-potential}
(a) the non-perturbative scalar potential for the complex position
modulus, $\phi^2_1$, of the D5-brane at
the origin in eq.(\ref{vacua2}). Its VEV determines the position of the D-brane along the
second 2-torus, keeping fixed the positions of the other branes. When the brane sits at the origin,
the system is in the supersymmetric vacuum and there is no scalar
potential for the Wilson line moduli of the magnetized
D9-branes. Outside this locus, however, magnetized D9-branes feel a potential. We have plotted in Figure \ref{junk-potential} (b) the non-perturbative scalar potential for the complex Wilson line modulus  of the magnetized D9-brane along the second 2-torus, $\xi^2_a$, for an arbitrary VEV of $\phi^2_1$. The minimum of this potential is at,
\begin{equation}
\xi^2_a=\frac14+\frac{\tau_2}{4} \quad \left(\textrm{mod} \quad \frac12,\ \frac{\tau_2}{2}\right) \label{minimax}
\end{equation}
Hence, while $\phi^2_1$ is rolling down towards the supersymmetric vacuum at the origin, the recombination moduli of the fractional magnetized branes feel a potential, inducing the spontaneous breaking of the gauge symmetry, eq.(\ref{u2break}). Fluctuations around the non-perturbative vacuum in this model will favor a gauge group $U(2)\times U(1)^{18}$.

\begin{figure}[!h]
\begin{center}
\includegraphics[width=6.5cm]{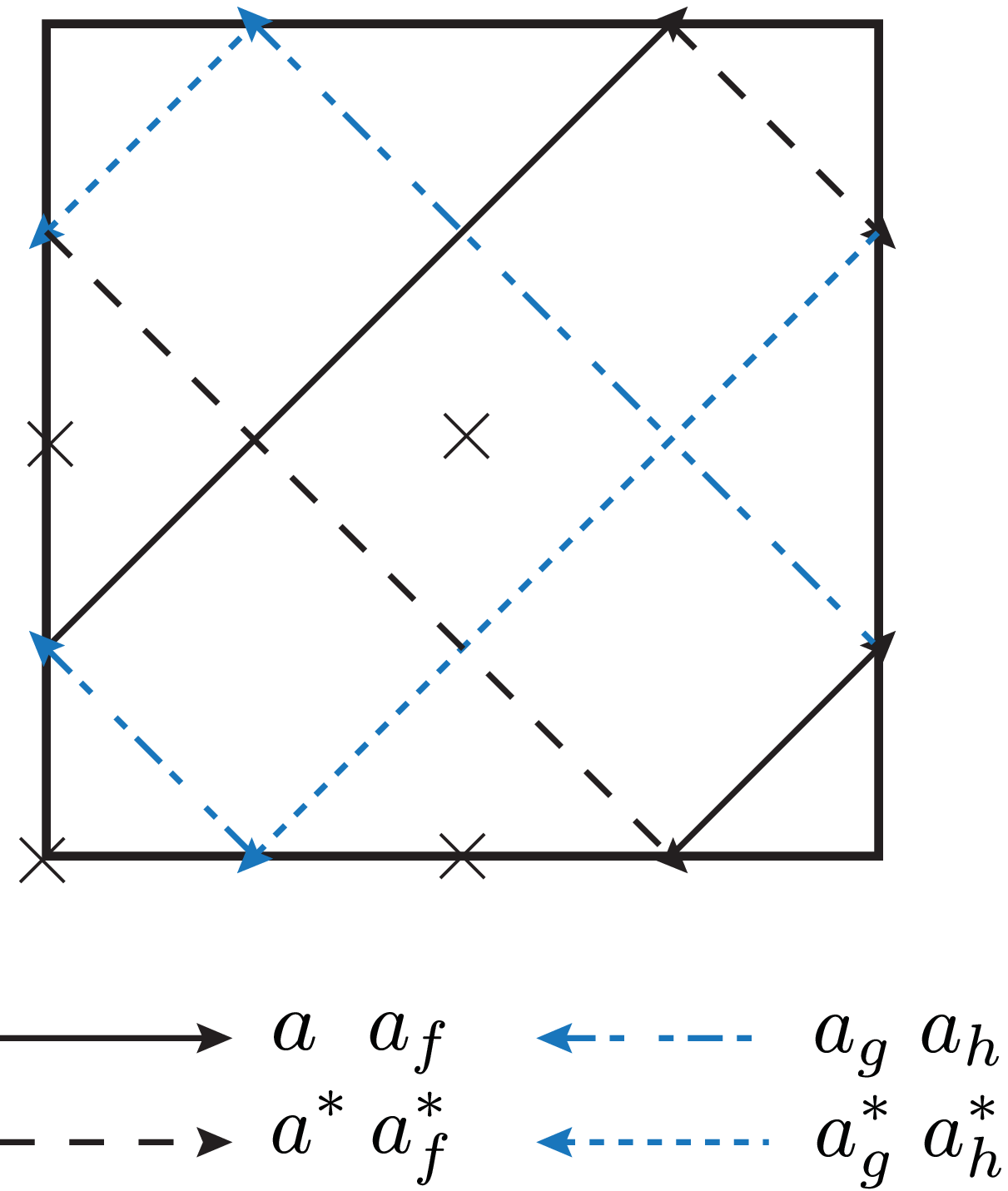}
\end{center}
\caption{\label{recom} Position of the recombined D6-brane and its images under the orientifold and orbifold actions on the second 2-torus, for the minimum of the scalar potential given in eq.(\ref{minimax}).}
\end{figure}

As already advanced, the minimum in eq.(\ref{minimax}) admits a nice geometric interpretation in the T-dual setup with intersecting D6-branes, corresponding to the equidistant position of the recombined D6-brane to the $\mathbb{Z}_2\times \mathbb{Z}_2$ singularities. At this locus, the non-perturbative forces which different instantons exert on the recombined brane are balanced. We have represented this situation in Figure \ref{recom}.

\subsection{A model with only magnetized branes}

The above results suggest that non-magnetized branes may play an important role in restoring the supersymmetry in models where linear superpotential couplings for some charged field are induced by rigid instantons. It is therefore natural to wonder about the existence of consistent models with only magnetized branes, but where linear superpotential couplings are still non-perturbatively generated. In this subsection we present one of such models, containing only magnetized D9-branes. Whereas in this case one cannot restore supersymmetry through the mechanism presented in the previous subsection, since there are no bulk non-magnetized D9 or D5 branes,  we will give arguments  based on the explicit form of the superpotential why supersymmetry is probably restored.

The model consists of four stacks of magnetized D9-branes with the following wrapping numbers
\begin{equation}
\begin{split}
1_p\ :\qquad &(m^i,n^i)= (-1,1)\otimes (-4,1) \otimes (1,2) \,,
\\
1_q\ : \qquad&(m^i,n^i)= (-1,1)\otimes (-4,1) \otimes (1,2) \,,
\\
2\ :\qquad &(m^i,n^i)= (-2,1)\otimes (-2,3) \otimes (1,2) \,,
\\
3\ : \qquad&(m^i,n^i)= (6,1) \otimes (-2,1) \otimes (-1,2) \,.
\end{split}\label{frac2}
\end{equation}
Branes $1_p$ and $2$ are of type $(p_a,q_\alpha)=(2,0)$ and $(p_a,q_\alpha)=(1,0)$ in eq.(\ref{cp}), whereas branes $1_q$ and $3$ are of type $(p_a,q_\alpha)=(0,2)$ and $(p_a,q_\alpha)=(0,1)$, respectively. The gauge group is $[U(2)\times U(1)]^2$. One may check that, among other fields, the massless spectrum of this model contains 30 degenerate fields transforming in the antisymmetric representation of $U(2)_{1_q}$, and that $E1$-brane instantons wrapping the third 2-torus have the right structure of zero modes to generate linear superpotential couplings for these fields.

Stacks $1_p$ and $1_q$ have the same magnetizations and therefore they can recombine, $1_p+1_q\to1$, allowing for continuous Wilson line deformations in the first 2-torus. The gauge group is broken to
\begin{equation}
U(2)\times U(1)\times U(1)\label{grupa}
\end{equation}
This recombination process is parameterized by VEV's of fields $\Phi_{1_p\bar1_q}$ and  $\tilde\Phi_{\bar1_p1_q}$, which are also in the massless spectrum. After recombination, the Wilson line modulus of the recombined brane in the first torus is related to the VEV of the bilinear $\xi^1_1\sim\langle\Phi_{1_p\bar1_q}\tilde\Phi_{\bar1_p1_q}\rangle$ which is contained in the adjoint representation of $U(2)$. We can follow the recombination from the point of view of the Higgs mechanism, as we did in subsection \ref{higgs}. The multiplicity of fields transforming in the antisymmetric and symmetric representations of the two $U(2)$ factors of the initial gauge group is given by eqs.(\ref{nafr}) and (\ref{nsfr}), respectively. However, the number of symmetric and antisymmetrics arising after recombination from bifundamentals is no longer given by eq.(\ref{r5}), since recombination in this model occurs along only one of the three 2-tori. We have instead,
\begin{equation}
N_{A}^{\rm bi} \ =  \ N_{S}^{\rm bi} \ = \ \frac{1}{4}\left( I_{aa^*} -  4I^1_{aa^*} + 4I^2_{aa^*} - 4I^3_{aa^*} \right)
\end{equation}
Hence, after recombination the total number of symmetrics and antisymmetrics of $U(2)$ is given respectively by,
\begin{align}
N_A \ & = \ 2 N_{A}^{\rm fr} + N_{A}^{\rm bi} \ =\ \frac{1}{2}\left( I_{aa^*} \ + \ \frac{1}{2} I_{aO}\ - \ 4I_{aa^*}^1 \right) \ ,\\
N_S \ & = \ 2 N_{S}^{\rm fr} + N_{S}^{\rm bi} \ =\ \frac{1}{2}\left( I_{aa^*} \ - \ \frac{1}{2} I_{aO}\ - \ 4I_{aa^*}^1 \right) \ .
\end{align}
We reproduce in Table \ref{tablare} the complete spectrum of this model which results after recombination. Similarly, we present in Table \ref{tablae1} the spectrum of instanton charged zero modes.
\begin{table}[htb!]
\begin{center}
\begin{tabular}{|c|c|c|}
\hline
Multiplicity & Representation & Field
\\[.5ex]
\hline
\hline
1 & $(4,1,1)_{(0,0,0)}$ & $\xi^1_{1}$
\\[.3ex]
\hline
66 & $(1,1,1)_{(2,0,0)}$ & $A^1$
\\[.3ex]
6& $(3,1,1)_{(2,0,0)}$ & $S^1$
\\[.3ex]
11 & $(1,1,1)_{(0,2,0)}$ & $S^2$
\\[.3ex]
15& $(1,1,1)_{(0,0,2)}$ & $S^3$
\\[.3ex]
\hline
10 & $(\bar2,1,\bar1)_{(-1,0,-1)}$ & $\Phi_{\bar1\bar3}$
\\[.3ex]
14 & $(2,1,\bar1)_{(1,0,-1)}$ & $\Phi_{1\bar3}$
\\[.3ex]
90 & $(2,1,1)_{(1,1,0)}$ & $\Phi_{12}$
\\[.3ex]
2 & $(2,\bar1,1)_{(1,-1,0)}$ & $\Phi_{1\bar2}$
\\[.3ex]
12 & $(1,\bar1,\bar1)_{(0,-1,-1)}$ & $\Phi_{\bar2\bar3}$
\\[.3ex]
48 & $(1,\bar1,1)_{(0,-1,1)}$ & $\Phi_{\bar23}$
\\[.3ex]
\hline
\end{tabular}
\end{center}
\caption{Charged massless spectrum of the model after recombination. The gauge group is broken to $U(2)\times U(1)\times U(1)$. We have also specified $U(1)$ charges as subscripts.\label{tablare}}
\end{table}
\begin{table}[htb!]
\begin{center}
\begin{tabular}{|c|c|c|}
\hline
CP choice & Multiplicity & Representation
\\[.5ex]
\hline
\hline
$1$ & $-$ & $-$
\\[.3ex]
\hline
$2$ &  $1$ & $(1,\bar1,1)_{(0,-1,0)}$
\\[.3ex]
$\ $ & $1$ & $(1,1,1)_{(0,0,1)}$
\\[.3ex]
\hline
$3$ & $1$ & $(\bar2,1,1)_{(-1,0,0)}$
\\[.3ex]
$\ $& $2$ & $(1,\bar1,1)_{(0,-1,0)}$
\\[.3ex]
\hline
$4$ & $1$ & $(\bar2,1,1)_{(-1,0,0)}$
\\[.3ex]
\hline
\end{tabular}
\end{center}
\caption{Charged zero-mode structure for $O(1)$ E1-brane instantons in the $U(2)\times U(1)\times U(1)$ model after recombination. Chan-Paton choices are defined in eq.(\ref{E1cp}).\label{tablae1}}
\end{table}
With this information at hand, we can then write the following superpotential,
\begin{equation}
W = W_p + W_2 + W_3 + W_4 \label{supno}
\end{equation}
where $W_p$ denotes the perturbative superpotential and $W_{2,3,4}$ are non-perturbative superpotentials corresponding to different choices of CP factor for the E1-brane instanton. Their explicit expression read,
\begin{align}
W_{p} &= \Phi_{\bar1\bar3}\left[S^3\left(\Phi_{1\bar3}+\Phi_{\bar1\bar3}S^1+\Phi_{\bar1\bar3}A^1+\Phi_{12}\Phi_{\bar2\bar3}\right)+\Phi_{\bar23}\left(\Phi_{12}+S^2\Phi_{1\bar2}\right)\right] \nonumber\\
&+ S^2\Phi_{\bar2\bar3}\left(\Phi_{\bar23}+S^3\Phi_{\bar2\bar3}\right)\label{wp}\\
W_{2}&=S^2\Phi_{\bar1\bar3}\left(\Phi_{1\bar2}+\Phi_{1\bar3}\Phi_{\bar23}\right)\\
W_{3} &=S^2\left(\Phi_{12}\Phi_{1\bar2}+S^3(\Phi_{1\bar3}\Phi_{1\bar3}+\Phi_{1\bar3}\Phi_{12}\Phi_{\bar2\bar3})\right) \\
W_{4}&=A^1(1+\Phi_{\bar1\bar3} S^3\Phi_{1\bar3}+\Phi_{\bar1\bar3}\Phi_{12}\Phi_{\bar23}+S^2\Phi_{\bar2\bar3}\Phi_{\bar23})+S^1(\Phi_{\bar1\bar3}S^3\Phi_{1\bar3}+\Phi_{\bar1\bar3}\Phi_{12}\Phi_{\bar23})\nonumber\\
&+\Phi_{1\bar3}\left(S^3\Phi_{1\bar3}+\Phi_{12}\Phi_{\bar23}+S^2\Phi_{1\bar2}\Phi_{\bar23}+S^3\Phi_{12}\Phi_{\bar2\bar3}\right)+S^2\Phi_{1\bar2}\Phi_{1\bar2} + \Phi_{12}\Phi_{12}\Phi_{\bar23}\Phi_{\bar2\bar3}\label{w4}
\end{align}
Note that we have set all coefficients equal to 1. In a complete analysis, like the one performed in previous sections, these coefficients are actually functions of the complex adjoint field which parameterize complex Wilson line deformations of the recombined brane along the first 2-torus.

Whereas a detailed analysis of the non-perturbative D-brane dynamics in this model is beyond the scope of this paper, we can make some observations which strongly suggest the existence of non-perturbative supersymmetric vacua also in this model.
First of all, we can cast the superpotential (\ref{supno}) into the form
\begin{equation}
W \ = \ X_i f_i (\phi_a) \ + \ g (\phi_a) \ , \label{is}
\end{equation}
where $X_i$, $i = 1 \cdots N_X$ are fields participating to supersymmetry breaking having R-charge two, whereas $\phi_a$,
$a = 1 \cdots N_{\phi}$, of R-charge zero, enforce $F_{X_i} \not=0$. This happens for $N_X > N_{\phi}$ and for $f_i$
generic functions of $\phi_a$. Notice that whereas $g (\phi_a)$ breaks R-symmetry, this does not change supersymmetry breaking, which
is entirely governed by $f_i (\phi_a)$.
Comparing (\ref{is}) with  (\ref{supno}), we find that contrary to the model presented in previous subsections, superpotential (\ref{supno}) does not preserve an R-symmetry.
From the expressions (\ref{wp})-(\ref{w4}) we may easily identify the fields which appear linearly in the superpotential. These are
$A^1$ and $S^1$, with total multiplicity 72. The multiplicity of the other fields coupling to them in the superpotential is much larger.  From Table \ref{tablare} and expressions (\ref{wp})-(\ref{w4}) we observe that the total number of fields which couple to these fields is 174.
Thus, it is expected that supersymmetry can be restored in this model by giving VEV's to these fields \cite{seiberg1,seiberg2}. This, in particular, involves processes of recombination between branes and their orientifold images, so that we expect the gauge group in eq.(\ref{grupa}) to be completely broken to a discrete subgroup at the non-perturbative supersymmetric vacuum.

\section{Global model with mass terms}
\label{globalmass}

In this Section we consider a model, first presented in Refs.\cite{cristinatorsion,mikerigid}, with gauge group $U(4)$. Hence, the second of the three conditions stated in the introduction for having linear superpotential terms is not satisfied, and quadratic superpotential terms are non-perturbatively generated instead. We can, however, easily extend the techniques of the previous sections to explicitly compute the mass matrix of this model, and to determine whether the mass terms vanish or not for particular values of the Wilson line moduli. The explicit computation of quadratic superpotential terms is particularly relevant in models where neutrino Majorana masses and/or supersymmetric $\mu$-terms are perturbatively forbidden by some global symmetry and have to be generated by non-perturbative effects \cite{blumencft1,blumencft2,ibanez}.

The model that we consider is given by a stack of branes in the bulk with magnetization,
\begin{equation}
(m^i, n^i) =(1,1) \otimes (1,1) \otimes (-1,1) \ . \label{mass01}
\end{equation}
The spectrum consists of 32 fields transforming in the antisymmetric representation of $U(4)$, which we shall denote $A^M$, $M=1,\dots,32$, and 3 complex adjoint fields which contain the Wilson line moduli of the brane for each of the three 2-tori.

One may easily verify that E1-brane instantons wrapping the third 2-torus in this model lead to mass terms for the antisymmetrics,
\begin{multline}
W_{\rm n.p.}= \sum_\alpha\int d^4xd^2\theta
d^4\eta \ e^{2 \pi i S_{E1}-\sum_M \left<\eta_i A^M_{ij}\eta_j\right>_{\alpha,{\rm disc}}}\\
=\sum_\alpha\int d^4xd^2\theta \ e^{2 \pi i \hat T_3}\sum_{M,K}\epsilon_{ijkl} \left< A^M_{ij}\right>_\alpha\left< A^K_{kl}\right>_\alpha\label{masa}
\end{multline}
where we sum over repeated $SU(4)$ indices $i,j,k,l$. The sum over $\alpha$ denotes a sum over all instanton configurations. As before, this is a sum over the 64 possible configurations for the discrete Wilson lines and positions of the instanton, eq.(\ref{E1pos}), and the 4 choices for the Chan-Paton charge eq.(\ref{E1cp}).

The disk amplitudes which appear in eq.(\ref{masa}) are given by,
\begin{equation}
\left< A^M_{ij}\right>_\alpha=\mathcal{D}^M_\alpha A^M_{ij}
\end{equation}
We can naturally split the index $M$ into $\{k,(i_1,i_2,i_3)\}$ with $k=1,\ldots,4$ labeling the four orbifold invariant combinations,
\begin{align}
\mathcal{D}^{1,(i_1,i_2,i_3)}_\alpha&=
\mathcal{D}^{(i_1,i_2,i_3)}_{aa^*E1}+
\mathcal{D}^{(i_1,i_2,i_3)}_{a_ga_g^*E1}+
\mathcal{D}^{(i_1,i_2,i_3)}_{a_fa_f^*E1}+ \mathcal{D}^{(i_1,i_2,i_3)}_{a_ha_h^*E1}\label{d1}\\
\mathcal{D}^{2,(i_1,i_2,i_3)}_\alpha&=
\mathcal{D}^{(i_1,i_2,i_3)}_{aa_g^*E1}+
\mathcal{D}^{(i_1,i_2,i_3)}_{a_ga^*E1}+
\mathcal{D}^{(i_1,i_2,i_3)}_{a_fa_h^*E1}+ \mathcal{D}^{(i_1,i_2,i_3)}_{a_ha_f^*E1}\\
\mathcal{D}^{3,(i_1,i_2,i_3)}_\alpha&=
\mathcal{D}^{(i_1,i_2,i_3)}_{aa_f^*E1}+
\mathcal{D}^{(i_1,i_2,i_3)}_{a_ga_h^*E1}+
\mathcal{D}^{(i_1,i_2,i_3)}_{a_fa^*E1}+ \mathcal{D}^{(i_1,i_2,i_3)}_{a_ha_g^*E1}\\
\mathcal{D}^{4,(i_1,i_2,i_3)}_\alpha&=
\mathcal{D}^{(i_1,i_2,i_3)}_{aa_h^*E1}+\mathcal{D}^{(i_1,i_2,i_3)}_{a_ga_f^*E1}
+ \mathcal{D}^{(i_1,i_2,i_3)}_{a_fa_g^*E1}+ \mathcal{D}^{(i_1,i_2,i_3)}_{a_ha^*E1}\label{d4}
\end{align}
where,
\begin{equation}
{ \mathcal{D}}^{(i_1,i_2,i_3)}_{ABE1}  =
\prod_{r=1}^3  e^{- i \pi |I^r_{AB}| \epsilon^r
  \epsilon^{r+3}} \vartheta\left[{\frac{i_r}{I^r_{AB}}+  \epsilon^r
    \atop  I_{AB}^r \epsilon^{r+3} }\right] (I^r_{BE1}\xi^{r}_A+ I^r_{E1A}\xi^{r}_B  \ ;\
\tau^r|I^r_{AB}|)
\end{equation}
The notation for the different brane images is specified in Appendix \ref{sym}. Indices
$i_r=0,1$, $r=1,2,3$ denote degenerate antisymmetrics localized at the various intersections in each of the three 2-tori $(|I^r_{aa^*}|=2)$.

We have then the following holomorphic superpotential
\begin{equation}
W_{\rm n.p.}=e^{2\pi i \hat T_3} \sum_{\alpha} \sum_{k,l=1}^4\sum_{i_1,i_2,i_3, \atop j_1,j_2,j_3=0}^1 \mathcal{D}^{k,(i_1,i_2,i_3)}_{\alpha}\mathcal{D}^{l,(j_1,j_2,j_3)}_{\alpha}\hat A^{k,(i_1,i_2,i_3)}\hat A^{l,(j_1,j_2,j_3)} \label{mass02}
\end{equation}
where holomorphic (`hatted') variables are defined as in Section \ref{sechol}.

Let us analyze the structure of these mass terms. For that, let us define
\begin{equation}
 g_r^{(i_r)} ({\xi_a^r,\tau_r},\epsilon_r,\epsilon_{r+3}) \ = \
e^{- 2 i \pi \epsilon_r \epsilon_{r+3}}
\left(\vartheta\left[{\frac{i_r}{2} + \epsilon_r \atop
 2 \epsilon_{r+3} }\right](2 \xi^r_a~;  2 \tau_r) +
\vartheta\left[{\frac{i_r}{2} + \epsilon_r \atop 2 \epsilon_{r+3}
  }\right](- 2 \xi^r_a~; 2\tau_r) \right) \ . \label{mass1}
\end{equation}
Then for the model under consideration, eq.(\ref{mass01}), it is possible to see that eqs.(\ref{d1})-(\ref{d4}) can be expressed as
\begin{align}
\mathcal{D}^{1,(i_1,i_2,i_3)}_{\alpha}&= g_1^{(i_1)} ({\xi_a^1,\tau_1}, \epsilon_1,\epsilon_4)  g_2^{(i_2)}
({\xi_a^2,\tau_2},\epsilon_2,\epsilon_5)  g_3^{(i_3)} ({0,\tau_3},\epsilon_3,\epsilon_6)
  \\
\mathcal{D}^{2,(i_1,i_2,i_3)}_{\alpha}&=  g_1^{(i_1)} ({\xi_a^1,\tau_1},\epsilon_1,\epsilon_4)  g_2^{(i_2)}
({0,\tau_2},\epsilon_2,\epsilon_5)  g_3^{(i_3)} ({\xi_a^3 ,\tau_3},\epsilon_3,\epsilon_6)
 \\
\mathcal{D}^{3,(i_1,i_2,i_3)}_{\alpha}&=  g_1^{(i_1)} ({0,\tau_1},\epsilon_1,\epsilon_4)  g_2^{(i_2)}
({\xi_a^2,\tau_2},\epsilon_2,\epsilon_5)  g_3^{(i_3)} ({\xi_a^3 ,\tau_3},\epsilon_3,\epsilon_6)
 \\
\mathcal{D}^{4,(i_1,i_2,i_3)}_{\alpha}&=  g_1^{(i_1)} ({0,\tau_1},\epsilon_1,\epsilon_4)  g_2^{(i_2)}
({0,\tau_2},\epsilon_2,\epsilon_5)  g_3^{(i_3)} ({0 ,\tau_3},\epsilon_3,\epsilon_6)
\end{align}
Thus, analogously with the linear terms studied in the preceding
section, mass terms for eight of the thirty-two
antisymmetric fields are non-zero and independent of the Wilson lines of the
magnetized bulk brane.

Moreover, note that superpotential (\ref{mass02}) can be cast into the form
\begin{equation}
W_{\rm n.p.} \ = \ e^{2\pi i \hat T_3} \ \sum_{\alpha} \ A_{\alpha} \ A_{\alpha} \ , \label{mass2}
\end{equation}
with
\begin{equation}
A_{\alpha} \ = \ \sum_k \sum_{i_1,i_2,i_3} \ \mathcal{D}^{k,(i_1,i_2,i_3)}_{\alpha} \ {\hat A}^{k,(i_1,i_2,i_3)} \ . \label{mass3}
\end{equation}
Therefore each instanton gives mass to one linear combination $A_{\alpha}$ of the antisymmetric fields. Since there are more instantons (64) than antisymmetric fields (32) one can in principle generate mass terms to enough independent linear combinations $A_{\alpha}$ to get masses for all antisymmetrics from instanton effects. In particular, the rank of the mass matrix is expected to increase.


\section{Conclusions}
\label{conclusions}

The main subject addressed in this paper is the fate of (linear) non-perturbative brane instabilities generated by instanton effects. These are usually invoked in various applications within QFT and String Theory, such as supersymmetry breaking, moduli stabilization and inflation. By using CFT techniques and transformation properties of closed string moduli and charged fields under monodromies of brane moduli, we were able to determine the consistent way of summing over discrete instanton configurations in the superpotential. This allowed us to find the structure of non-perturbative vacua of such models, which in the examples we study turn out to be supersymmetric and interpreted in terms of D-brane recombination, gauge symmetry breaking and partial open string moduli stabilization.

Thus, the results of this work suggest that non-perturbative vacuum destabilization in String Theory compactifications generically leads to spontaneous breaking of gauge symmetries rather than to supersymmetry breaking. Whereas we cannot exclude the latter, it remains to be seen if there are models where the non-perturbative D-brane recombination that we have described can give rise to supersymmetry breaking vacua, as advocated in \cite{gaugemed}. Note that in this case, even if the open string sector were to admit supersymmetry breaking, there might still be runaway directions in the closed string sector.

Related to this, we have also pointed out the importance of global aspects when analyzing non-perturbative effects. D-branes generically intersect rigid instantons in various singularities (contractible cycles) of the compact space. The local non-perturbative dynamics induced on the branes is usually the result of forces exerted by instantons located at different positions in the compact space. This is particularly relevant for toroidal models, where intersecting/magnetized branes, being fully rotated in the three internal tori, necessarily intersect rigid instantons in {\it all singularities} of the compact space.
Non-standard geometries, such as highly-warped spaces \cite{holo,holo2}, are therefore required in order for a local description of non-perturbative effects to be reliable.

There are various potential applications of the processes of brane recombination and gauge symmetry breaking which we have presented. These include the breaking of gauge symmetries of hidden sector(s) or the reduction of the rank of the gauge group, leading to open string moduli stabilization. In this sense, the tendency of brane recombinations to erase instanton effects raises further question marks about explicit implementations of the KKLT scenario of moduli stabilization \cite{kklt}.

Another interesting application of the non-perturbative dynamics discussed in this work, and more precisely of the motion of D-branes generated by instantonic forces, is to D-brane inflationary models. Indeed, under some assumptions, this motion can be slow enough to induce an inflationary universe. In this context, notice that relative phases between different instanton contributions  are crucial in obtaining consistent global superpotentials. We have determined these phase factors by exploiting invariance of the superpotential under monodromies of the D-brane scalars (Wilson line and geometric moduli), but it would be also desirable to compute them by more direct means, such as the localization techniques of \cite{nikita}, which have been recently extended to String Theory instantons \cite{higher4,lerda}. The role of relative phase factors in inflationary models with several condensates or instanton effects awaits more dedicated studies.

Finally, the fact of having generically various instantons which contribute at the same order to a given superpotential coupling can also be relevant in a broader context. As we argued in Section 5, the summation over instanton positions increases the rank of the mass matrix. Similar considerations may apply to other couplings, such as Yukawa couplings. Hence, global aspects can also have a potential impact on the phenomenological features of quasi-realistic brane models.


\section*{Acknowledgments}
{We thank F. Marchesano and E. Palti for useful discussions and
  comments. P.G.C. and E.D. thank the GGI Institute in Florence,
  P.G.C. thanks Ecole Polytechnique and E.D. thanks
the Aspen Center for Physics for hospitality during the completion of this work.  Work partially supported by
the European ERC Advanced Grant 226371 MassTeV, by the CNRS
PICS no. 3747 and 4172, CNRS grant ANR-05-BLAN-0079-02 and the grant CNCSIS `Idei' 454/2009, ID-44.}

\appendix

\section{Computation of one-loop amplitudes}
\label{thresholds}

In this appendix we present the computation of the one-loop amplitudes involving the fractional $E1$-brane instantons wrapping the third 2-torus and non-magnetized bulk D9-branes and D5-branes, contributing to the determinant of the one-loop fluctuations \cite{abel}. For simplicity we only explicitly consider bulk D5-branes wrapping the first 2-torus. Generalization to bulk D5-branes wrapping the second 2-torus is straightforward. Similar results could have been obtained by computing the gauge threshold corrections of a D5-brane wrapping the third torus \cite{berg,maldacena,lust} (see also \cite{stieberg,blumenthres,gabi}).

\subsection*{M\"obius amplitude}
The M\"obius amplitude for $E1$-brane instantons wrapping the third 2-torus is the following
\begin{equation}
\begin{split}
\mathcal{M}_{{\rm E}1} & = \frac{1}{8}
D_{h;o}\int_0^{\infty}\frac{dt}{t}\left(\frac{2\hat\eta}{\hat\vartheta_2}\right)^2\Bigg\lbrace
-\hat{T}_{oo}^*\, W_1W_2P_3
\\
& +\left[-\hat{T}_{og}^* \, W_1 -\hat{T}_{of}^*\, W_2-
\hat{T}_{oh}^*\, P_3\right]
\left(\frac{2\hat\eta}{\hat\vartheta_2}\right)^2\Bigg\rbrace \,.
\\
\end{split}\label{mob}
\end{equation}
where $W_i$ and $P_i$ represent the standard winding and momentum lattice sums, and
\begin{align}
\hat{T}_{oo}^*&=\frac{1}{2\hat\eta^4}\sum_{\alpha\beta}c_{\alpha\beta}\,
\hat\vartheta \left[ {\alpha \atop \beta+1/2}\right] \hat\vartheta \left[ {\alpha
\atop \beta-1/2}\right] \hat\vartheta^2 \left[ {\alpha \atop
\beta}\right]=0\\
\hat{T}_{ok}^*&=\frac{1}{2\hat\eta^4}\sum_{\alpha\beta}c_{\alpha\beta}\,
\hat\vartheta^2 \left[ {\alpha \atop \beta+1/2}\right] \hat\vartheta^2 \left[
{\alpha \atop \beta-1/2}\right]
=-\left(\frac{\hat\vartheta_2}{\hat\eta}\right)^4
\end{align}
with $k=g,f,h$,
$c_{\alpha\beta}=(-1)^{2(\alpha+\beta+2\alpha\beta)}$ and $\alpha,\beta$ taking values $0$ or $1/2$. Hatted functions indicate that the parameter is the standard M\"obius parameter, $\frac{it}{2}+\frac12$. For more details in the conventions we use see \cite{dudas,sagnotti}. From the evaluation of eq.(\ref{mob}) we obtain
\begin{equation}
\mathcal{M}_{{\rm E}1}  = 2
D_{h;o}\int_0^{\infty}\frac{dt}{t}\left(W_1+W_2+P_3\right)
\end{equation}
with $D_{h;o}$ the untwisted Chan-Paton charge of the instanton, which for our purposes will be equal to one. Notice, in particular, that since the instanton is placed on top of the orientifold planes, the M\"obius amplitude is independent of the position of the instanton.

The integral can be evaluated by introducing an ultraviolet cut-off, $\Lambda$, and an infrared regulator,
$\mu$, using the formulas in \cite{abd,ghilencea1,ghilencea2}
\begin{align}
&\int_{1/\Lambda^2}^{\infty}\frac{dt}{t}P_3e^{-2\pi\mu t}=\textrm{Im }
T_3\Lambda^2-\ln\left(8\pi^3\mu\, \textrm{Im } T_3\, \textrm{Im }\tau_3\,
|\eta(\tau_3)|^4\right)\\
&\int_{1/\Lambda^2}^{\infty}\frac{dt}{t}W_{1,2}e^{-2\pi\mu
t}=\frac{\Lambda^2}{2\textrm{Im } T_{1,2}}-\ln\left(\frac{4\pi^3\mu\, \textrm{Im }
\tau_{1,2}}{\textrm{Im } T_{1,2}}\, |\eta(\tau_{1,2})|^4\right)
\end{align}

\subsection*{Annulus amplitudes}
The annulus amplitude between a $E1$-brane instanton wrapping the third 2-torus and a
non-magnetized bulk D9-brane with a Wilson line also in the third torus and untwisted Chan-Paton charge $2(n_1+n_2)$ reads
\begin{multline}
\mathcal{A}_{E1-D9}=
  \frac{1}{8}\int_0^{\infty}\frac{dt}{t}\left(\frac{\eta}{\vartheta_4}\right)^2
\Bigg\lbrace D_{h;o}(n_1+n_2)\frac{\eta}{\vartheta\left[ {\alpha \atop
\beta}\right]}\frac{\vartheta^4 \left[ {\alpha+\frac12 \atop
\beta}\right]\vartheta \left[ {\alpha \atop
\beta}\right]}{\eta^5}\left(\frac{\eta}{\vartheta_4}\right)^2\\
  +D_{h;h}(n_1-n_2)\frac{\eta}{\vartheta\left[ {\alpha \atop
\beta}\right]}\frac{\vartheta^2 \left[ {\alpha+\frac12 \atop
\beta}\right]\vartheta \left[ {\alpha+\frac12 \atop \beta+\frac12}\right]\vartheta
\left[ {\alpha+\frac12 \atop \beta-\frac12}\right]\vartheta\left[ {\alpha \atop
\beta}\right]}{\eta^5}\left(\frac{\eta}{\vartheta_3}\right)^2\Bigg\rbrace \\
\times \
\left(P_{\vec m^3-\vec \theta_{\rm E1}+\vec \theta_{\rm D9}}+P_{\vec m^3+\vec \theta_{\rm E1}-\vec \theta_{\rm D9}}\right)
\end{multline}
Using the identity
\begin{equation}
\sum_{\alpha\beta}c_{\alpha\beta}\, \vartheta^4 \left[ {\alpha+1/2
\atop \beta}\right]=-2\vartheta_4^4
\end{equation}
and $n_1=n_2=N_{D9}$ for $N_{D9}$ bulk D9-branes, we get
\begin{equation}
\mathcal{A}_{E1_3-D9}=-\frac{N_{D9}}{2}\int_0^{\infty}\frac{dt}{t}\left(P_{\vec m^3-\vec \theta_{\rm E1}+\vec \theta_{\rm D9}}+P_{\vec m^3+\vec \theta_{\rm E1}-\vec \theta_{\rm D9}}\right)
\end{equation}
In this case the integral is finite in the infrared (for $\vec \theta_{\rm E1}-\vec \theta_{\rm D9}\neq 0$) and only regularization of the ultraviolet is required \cite{abd,ghilencea1,ghilencea2}
\begin{equation}
\int_{1/\Lambda^2}^{\infty}\frac{dt}{t}P_{\vec m^3\pm \vec \theta_{\rm E1}\mp \vec \theta_{\rm D9}}=\textrm{Im }
T_3\Lambda^2-\ln\left|
\frac{\vartheta_1(\xi^3_{D9E1};\tau_3)}{\eta(\tau_3)}\right|^2 +2\pi\frac{\left(\textrm{Im }
\xi^3_{D9E1}\right)^2}{\textrm{Im } \tau_3}
\end{equation}
where $\xi^3_{D9E1}=\xi^3_q-\xi^3_{\rm E1}$.

Similarly, annulus amplitudes between the $E1$-brane instanton and
non-magnetized $D5$-branes wrapping the first 2-torus, with untwisted Chan-Paton charge $2(d_1+d_2)$ and a Wilson line in the second torus
reads
\begin{multline}
\mathcal{A}_{E1-D5_1}=
  \frac{1}{8}\int_0^{\infty}\frac{dt}{t}\left(\frac{\eta}{\vartheta_4}\right)^2
\Bigg\lbrace D_{h;o}(d_1+d_2)\frac{\eta}{\vartheta\left[ {\alpha \atop
\beta}\right]}\frac{\vartheta^3 \left[ {\alpha+\frac12 \atop
\beta}\right]\vartheta \left[ {\alpha \atop \beta}\right]\vartheta \left[
{\alpha+\frac12 \atop
\beta}\right]}{\eta^5}\left(\frac{\eta}{\vartheta_4}\right)^2\\
  + D_{h;f}(d_1-d_2)\frac{\eta}{\vartheta\left[ {\alpha \atop
\beta}\right]}\frac{\vartheta^2 \left[ {\alpha+\frac12 \atop
\beta}\right]\vartheta \left[ {\alpha+\frac12 \atop \beta+\frac12}\right]\vartheta
\left[ {\alpha \atop \beta}\right]\vartheta\left[ {\alpha+\frac12 \atop
\beta-\frac12}\right]}{\eta^5}\left(\frac{\eta}{\vartheta_3}\right)^2\Bigg\rbrace\\
\times \
\left(W_{\vec n^2-\vec \varepsilon_{\rm E1}+\vec \varepsilon_{\rm D5}}+W_{\vec n^2+\vec \varepsilon_{\rm E1}-\vec \varepsilon_{\rm D5}}\right)
\end{multline}
Using the same identity as before and $d_1=d_2=N_{D5_1}$ for $N_{D5_1}$ bulk D5-branes one gets
\begin{equation}
\mathcal{A}_{E1-D5_1}=-\frac{N_{D5_1}}{2}\int_0^{\infty}\frac{dt}{t}\left(W_{\vec n^2-\vec \varepsilon_{\rm E1}+\vec \varepsilon_{\rm D5}}+W_{\vec n^2+\vec \varepsilon_{\rm E1}-\vec \varepsilon_{\rm D5}}\right)
\end{equation}
Then we can evaluate the
integral as \cite{abd,ghilencea1,ghilencea2}
\begin{equation}
\int_{1/\Lambda^2}^{\infty}\frac{dt}{t}W_{\vec n^2\mp\vec \varepsilon_{\rm E1}\pm\vec \varepsilon_{\rm D5}}=\frac{\Lambda^2}{\textrm{Im } T_2}-\ln\left|
\frac{\vartheta_1(\phi^2_{E1D5};\tau_2)}{\eta(\tau_2)}\right|^2 +2\pi\frac{\left(\textrm{Im }
\phi^2_{E1D5}\right)^2}{\textrm{Im } \tau_2}
\end{equation}
where $\phi^2_{E1D5}=\phi^2_q-\phi^2_{D5}$.

\subsection*{One-loop determinant}

Putting everything together one obtains
\begin{multline}
- \mathcal{A}_{E1-D9} - \mathcal{A}_{E1-D5_1} - \mathcal{M}_{E1}\\
  = \int_0^{\infty}\frac{dt}{t}\Big\lbrace \frac{N_{D9}}{2}\left(P_{\vec m^3-\vec \theta_{\rm E1}+\vec \theta_{\rm D9}}+P_{\vec m^3+\vec \theta_{\rm E1}-\vec \theta_{\rm D9}}\right)+
\frac{N_{D5_1}}{2}\left(W_{\vec n^2-\vec \varepsilon_{\rm E1}+\vec \varepsilon_{\rm D5}}+W_{\vec n^2+\vec \varepsilon_{\rm E1}-\vec \varepsilon_{\rm D5}}\right)\\
-2(W_1+W_2+P_3)\Big\rbrace
  = N_{D9}\left[-\ln\left|
\frac{\vartheta_1(\xi^3_{E1D9};\tau_3)}{\eta(\tau_3)}\right|^2 +2\pi\frac{\left(\textrm{Im }
\xi^3_{E1D9}\right)^2}{\textrm{Im } \tau_3}\right]\\
  +N_{D5_1}\left[-\ln\left|
\frac{\vartheta_1(\phi^2_{E1D5};\tau_2)}{\eta(\tau_2)}\right|^2 +2\pi\frac{\left(\textrm{Im }
\phi^2_{E1D5}\right)^2}{\textrm{Im } \tau_2}\right]-\Lambda^2\left[\frac{1}{\textrm{Im } T_{1}}
+\frac{1-N_{D5_1}}{\textrm{Im } T_{2}}+(2-N_{D9})\textrm{Im } T_3\right]\\+2\ln\left[\left(\frac{4\pi^3\mu\, \textrm{Im }
\tau_{1}}{\textrm{Im } T_{1}}\, |\eta(\tau_{1})|^4\right)\left(\frac{4\pi^3\mu\, \textrm{Im }
\tau_{2}}{\textrm{Im } T_{2}}\, |\eta(\tau_{2})|^4\right)\left(8\pi^3\mu\, \textrm{Im } T_3\, \textrm{Im }\tau_3\,
|\eta(\tau_3)|^4\right)\right]
\end{multline}\label{unnormalized}
This expression should be interpreted with a gauge threshold. The normalization can be fixed by imposing that the chiral fermions becoming massless when $\xi^3_{E1D9}=0$ or $\phi^2_{E1D5}=0$, contribute in the right amount to the coefficient of the $\beta$-function, given by the field theory formula,
\begin{equation}
\frac{4\pi}{g^2}=\frac{4\pi}{g_0^2}+\frac{1}{2}\left(3C_2(G)-\sum_R
T(R)\right)\ln\frac{\mu}{M_s}
\end{equation}
The gauge threshold will  be then identified with the real part of the logarithm of the quantity which enters in $\langle F_A\rangle$, obtaining
\begin{multline}
\textrm{exp}\left(\mathcal{A}_{E1_\alpha}+\mathcal{M}_{E1_\alpha}\right)
= {\cal N} \left[\eta(\tau_1)
\eta(\tau_2)^{1+2N_{D5_1}}\eta(\tau_3)^{1+2N_{D9}}\right]^{-1}\times\\
\prod_{k=1}^{N_{D9}}\left(\textrm{exp}\left[\frac{2\pi i (\xi^3_{k}
      \textrm{Im }\xi^3_{k} +\xi^3_{E1_{\alpha}} \textrm{Im
      }\xi^3_{E1_{\alpha}})}{\textrm{Im }\tau_3}\right]
\vartheta\left[{\frac{1}{2} \atop \frac{1}{2}}\right](\xi^3_k
+\xi^3_{E1_\alpha} ; \tau_3)  \vartheta\left[{\frac{1}{2} \atop \frac{1}{2}}\right](\xi^3_k
-\xi^3_{E1_\alpha} ; \tau_3) \right)  \times \\
\prod_{q=1}^{N_{D5_1}}
\left(\textrm{exp}\left[\frac{2\pi i (\phi^2_{q}
      \textrm{Im }\phi^2_{q} +\phi^2_{E1_{\alpha}} \textrm{Im
      }\phi^2_{E1_{\alpha}})}{\textrm{Im }\tau_2}\right]
\vartheta\left[{\frac{1}{2} \atop \frac{1}{2}}\right](\phi^2_q
+\phi^2_{E1_\alpha} ; \tau_2)  \vartheta\left[{\frac{1}{2} \atop
    \frac{1}{2}}\right] (\phi^2_q -\phi^2_{E1_\alpha} ; \tau_2) \right)
\label{deter}
\end{multline}
where ${\cal N}$ is an irrelevant normalization factor.

\section{SUSY, Tadpoles and K-Theory Constraints}
\label{KTheory}

Compactifications with discrete torsion involve at least one exotic O-plane, with both positive tension and charge. In order to cancel the RR tadpoles ones has therefore to include also a suitable number of D5-branes and magnetized D9-branes, with constant magnetic fields $H_i$ along the $i$-th $T^2$,
\begin{equation}
H_i=\frac{m^i}{v_i n^i}\qquad i=1,2,3
\end{equation}
where $v_i$ is the volume of the $i$-th $T^2$.  Moreover, in order for $\mathcal{N}=1$ supersymmetry to be preserved in four dimensions,  the magnetic fields for the D9 branes must satisfy the following conditions,
\begin{eqnarray}
H_1+H_2+H_3=H_1H_2H_3 \nonumber\\
H_1H_2+H_1H_3+H_2H_3 \leq 1.
\end{eqnarray}
The first condition guarantees that an $\mathcal{N}=1$ supersymmetry is preserved whereas the latter condition guarantees that the {\it same} $\mathcal{N}=1$ supersymmetry is preserved by all stacks.

We shall focus on the specific choice of discrete torsion discussed in Section \ref{stringy}, namely $(\epsilon_g,\epsilon_f,\epsilon_h) = (+,+,-)$. The RR tadpole cancelation conditions then read
\begin{equation}
\begin{split}
\sum_A N_A \, n^1_{A}n^2_{A}n^3_{A} &=16 \,,
\\
\sum_A N_A \, n^1_{A}m^2_{A}m^3_{A} &=-16\, ,
\\
\sum_A N_A \, m^1_{A}n^2_{A}m^3_{A} &=-16 \, ,
\\
\sum_A N_A \, m^1_{A}m^2_{A}n^3_{A} &=16 \, ,
\\
\end{split} \label{untwisted}
\end{equation}
where the sums run over all the stacks of D5-branes and D9-branes. The unusual sign of last equation is due to the exotic O5-plane wrapping the third 2-torus.

Due to the coupling to twisted fields, fractional branes have also to satisfy twisted tadpole conditions for each of the $3\times 16$ fixed points. For the choice of Chan-Paton charges (\ref{cp}) these conditions read,
\begin{equation}
\begin{split}
\sum_a p_a \, m^1_{a}\epsilon_l^{(a),g}+\sum_{\alpha} q_{\alpha}\, m^1_{\alpha}\epsilon_l^{(\alpha),g} &= \ 0 \,,
\\
\sum_a p_a \, m^2_{a}\epsilon_l^{(a),f}-\sum_{\alpha} q_{\alpha} \, m^2_{\alpha}\epsilon_l^{(\alpha),f} & = \ 0 \,,
\\
\sum_a p_a \, n^3_{a}\epsilon_l^{(a),h}-\sum_{\alpha} q_{\alpha} \, n^3_{\alpha}\epsilon_l^{(\alpha),h} & = \ 0 \,,
\\
\end{split} \label{twisted}
\end{equation}
where
$\epsilon_l^{(A),g}$ is equal to $1$ if brane $A$ passes through the $l$-th point fixed under the action of the $g$ generator, and equals zero otherwise.  In addition to the tadpole constraints, a model must also satisfy K-theory constraints~\cite{angelk}.

In addition to tadpole constraints, self-consistent string models must also satisfy K-theory constraints.  We shall follow a similar approach to that of Refs. \cite{angelk,MS, blumenrigid}, which is to consider all of the possible D-brane probes that can be introduced with a $USp(2) \simeq SU(2)$ gauge group in their world-volume.  The constraints can then be derived by requiring that the number of fundamentals of $SU(2)$ arising from the intersections of the probe branes and the physical branes must be even.  For the choice of discrete torsion considered in this paper, we have the standard K-theory constraints,
\begin{eqnarray}
 \sum_A P_A~m^1_{A} m^2_{A} m^3_{A} &\in& 8 \mathbb{Z} \nonumber\\
 \sum_A P_A~m^1_{A} n^2_{A} n^3_{A} &\in& 8 \mathbb{Z} \\
 \sum_A P_A~n^1_{A} m^2_{A} n^3_{A} &\in& 8 \mathbb{Z}. \nonumber
\end{eqnarray}
In addition, because of the discrete torsion in the third torus, we also an additional set of constraints namely,
{\small \begin{eqnarray}
\sum_a p_a (n^1_a n^2_a m^3_a - S^a_g n^1_a -S^a_f n^2_a+S^a_h m^3_a) + \sum_\alpha q_\alpha (n^1_\alpha n^2_\alpha m^3_\alpha
- S^\alpha_g n^1_\alpha +S^\alpha_f n^2_\alpha-S^\alpha_h m^3_\alpha) \in 8 \mathbb{Z}\nonumber \\
\sum_a p_a (n^1_a n^2_a m^3_a - S^a_g n^1_a + S^a_f n^2_a -S^a_h m^3_a) + \sum_\alpha q_\alpha (n^1_\alpha n^2_\alpha m^3_\alpha
- S^\alpha_g n^1_\alpha -S^\alpha_f n^2_\alpha+S^\alpha_h m^3_\alpha) \in 8 \mathbb{Z}\nonumber \\
\sum_a p_a (n^1_a n^2_a m^3_a + S^a_g n^1_a -S^a_f n^2_a-S^a_h m^3_a) + \sum_\alpha q_\alpha (n^1_\alpha n^2_\alpha m^3_\alpha
+ S^\alpha_g n^1_\alpha +S^\alpha_f n^2_\alpha+S^\alpha_h m^3_\alpha) \in 8 \mathbb{Z}\nonumber \\
\sum_a p_a (n^1_a n^2_a m^3_a + S^a_g n^1_a + S^a_f n^2_a + S^a_h m^3_a) + \sum_\alpha q_\alpha (n^1_\alpha n^2_\alpha m^3_\alpha
+ S^\alpha_g n^1_\alpha - S^\alpha_f n^2_\alpha-S^\alpha_h m^3_\alpha) \in 8 \mathbb{Z}\nonumber\\
\label{D53}
\end{eqnarray}}
where $S^i_j$ refers to the number of shared fixed points of the $j^{\rm th}$ orbifold action between the $i^{\rm th}$ physical brane and a $D5$ probe brane wrapping the third 2-torus.  So for example, a bulk physical brane would have $S_j = 0$.

The constraints in eq.(\ref{D53}) must also be supplemented by considering the effects of discrete Wilson lines on the $D5$ probe branes wrapping the third 2-torus.  The net effects of these Wilson lines will be to change which fixed points the probe branes pass through.  This has the potential to change the $S^i_j$ terms.  For the setup considered in this paper, the only $S^i_j$ terms that can possibly be affected by discrete Wilson lines are those equal to two or four (with those equal to two being affected by a discrete Wilson line in one torus and those equal to four being affected by discrete Wilson lines in two torii).  In order to verify that a collection of branes satisfies the K-theory constraints, one may just verify that eq.(\ref{D53}) is satisfied for every choice of discrete Wilson line.

The requirement that eq. (\ref{D53}) be satisfied even when discrete Wilson lines are considered may seem to onerous to satisfy for any set of non-trivial K-theory charges.  However, the  twisted tadpole constraints guarantee that no fixed point is only intersected by only a single physical brane.  This implies that only {\it collections} of $S^i_j$ will change simultaneously when these discrete Wilson lines are considered.  As such, even in the presence of non-trivial K-theory charges collections of branes can still be found which satisfy the K-theory constraints.

It is interesting to note that as the most stringent K-theory constraints come from probe branes with the same wrapping numbers as the $E1$ instantons that these constraints also constrain the possible operators that are instantonically generated.  We note that eq.(\ref{D53}) implies that there can be no net odd charged operators generated by instantons in these models.

\section{F-terms for the model of Section \ref{simple}}
\label{sym}

In this Appendix we perform the symmetrization of the disc contributions to the superpotential with respect to all the orbifold operations, for the example in Section \ref{simple}. For that, one has to take into account that the orbifold generators act on the magnetization numbers, on the position and Wilson line moduli and on the vector $\vec i$ denoting the intersection on which the corresponding chiral field is localized.

Let us consider the recombined brane $a$ and its images under the orbifold and orientifold operations,
\begin{align}
a :  & \ (m^1,n^1)\otimes (m^2,n^2)\otimes (m^3,n^3)\label{images}&
\quad a^* :  & \ (-m^1,n^1)\otimes (-m^2,n^2)\otimes (-m^3,n^3)\nonumber\\
a_g :  & \ (m^1,n^1)\otimes (-m^2,-n^2)\otimes (-m^3,-n^3)\nonumber&
\quad a_g^* :  & \ (-m^1,n^1)\otimes (m^2,-n^2)\otimes (m^3,-n^3)\nonumber\\
a_f :  & \ (-m^1,-n^1)\otimes (m^2,n^2)\otimes (-m^3,-n^3)\nonumber&
\quad a_f^* :  & \ (m^1,-n^1)\otimes (-m^2,n^2)\otimes (m^3,-n^3)\nonumber\\
a_h :  & \ (-m^1,-n^1)\otimes (-m^2,-n^2)\otimes (m^3,n^3)\nonumber&
\quad a_h^* :  & \ (m^1,-n^1)\otimes (m^2,-n^2)\otimes (-m^3,n^3)\nonumber
\end{align}
We can express the holomorphic part of the disc amplitude by considering invariant amplitudes between the brane and different images. For that it is convenient to define the functions,
\begin{align}
\psi^{1,(i_1,i_1',i_2,i_3)}&=
\mathcal{F}^{(i_1,i_2,i_3)}_{aa^*E1}+
\mathcal{F}^{(i_1,i_2,i_3)}_{a_ga_g^*E1}+
\mathcal{F}^{(i_1',i_2,i_3)}_{a_fa_f^*E1}+ \mathcal{F}^{(i_1',i_2,i_3)}_{a_ha_h^*E1}\\
\psi^{2,(i_1,i_1',i_2,i_3)}&=
\mathcal{F}^{(i_1,i_2,i_3)}_{aa_g^*E1}+
\mathcal{F}^{(i_1,i_2,i_3)}_{a_ga^*E1}+
\mathcal{F}^{(i_1',i_2,i_3)}_{a_fa_h^*E1}+ \mathcal{F}^{(i_1',i_2,i_3)}_{a_ha_f^*E1}\\
\psi^{3,(i_1,i_1',i_2,i_3)}&=
\mathcal{F}^{(i_1,i_2,i_3)}_{aa_f^*E1}+
\mathcal{F}^{(i_1,i_2,i_3)}_{a_ga_h^*E1}+
\mathcal{F}^{(i_1',i_2,i_3)}_{a_fa^*E1}+ \mathcal{F}^{(i_1',i_2,i_3)}_{a_ha_g^*E1}\\
\psi^{4,(i_1,i_1',i_2,i_3)}&=
\mathcal{F}^{(i_1,i_2,i_3)}_{aa_h^*E1}+\mathcal{F}^{(i_1,i_2,i_3)}_{a_ga_f^*E1}
+ \mathcal{F}^{(i_1',i_2,i_3)}_{a_fa_g^*E1}+ \mathcal{F}^{(i_1',i_2,i_3)}_{a_ha^*E1}
\end{align}
where
\begin{multline}
{ \mathcal{F}}^{(i_1,i_2,i_3)}_{ABE1}  =   \ \
\left(\prod_{r=1}^3  \sum_{\epsilon^r,\epsilon^{r+3}=
  0,1/2}  e^{- 2 \pi i [N_{D9}(\epsilon^6+ \epsilon^3
\epsilon^6 )+ N_{D5_1} (\epsilon^5+\epsilon^2
\epsilon^5 ) ]}  e^{2 \pi i (T_3+ M_{\alpha})- i \pi |I^r_{AB}| \epsilon^r
  \epsilon^{r+3}}  \right. \\
\left. \vartheta\left[{\frac{i_r}{I^r_{AB}}+  \epsilon^r
    \atop  I_{AB}^r \epsilon^{r+3} }\right] (I^r_{BE1}\xi^r_A+ I^r_{E1A}\xi^r_B \ ;\
\tau_r|I^r_{AB}|) \right)   \\ \left.
\times \
\prod_{k=1}^{N_{D9}}
\vartheta^2\left[{\frac{1}{2} + \epsilon^3 \atop \frac{1}{2}
    + \epsilon^6 }
\right](\xi^3_k ; \tau_3)  \times \prod_{q=1}^{N_{D5_1}}
\vartheta^2 \left[{\frac{1}{2} + \epsilon^2 \atop
    \frac{1}{2} +  \epsilon^5 }\right] (\phi^2_q ; \tau_2)
\right.   \label{Fterms}
\end{multline}
where $i_1=0\ldots 3$, $i_{2,3}=0,1$, denote the intersections in each of
the three 2-torus, from left to right.
Let us define in what follows the functions
\begin{eqnarray}
&& f_1^{(i_1i'_1)} ({\xi_a^1,\tau_1}) = \sum_{\epsilon_1,\epsilon_4}
e ^{- 4 i \pi \epsilon_1 \epsilon_4}
\left(\vartheta\left[{\frac{i_1}{4} + \epsilon_1 \atop
 4 \epsilon_4 }\right](2 \xi^1_a ;  4\tau_1) +
\vartheta\left[{\frac{i_1'}{4} + \epsilon_1 \atop 4 \epsilon_4
  }\right](- 2 \xi^1_a ; 4\tau_1)\right) \ , \nonumber \\
&& f_2^{(i_2)} ({\xi_a^2, \phi^2_q, \tau_2, N_{D5_1}}) = \sum_{\epsilon_2,\epsilon_5}
e ^{- 2 i \pi [\epsilon_2 \epsilon_5 + N_{D5_1} (\epsilon_5 +
  \epsilon_2 \epsilon_5)] } \nonumber \times \\
&& \left(\vartheta\left[{\frac{i_2}{2} + \epsilon_2 \atop
 2 \epsilon_5 }\right](2 \xi^2_a ;  2\tau_2) +
\vartheta\left[{\frac{i_2}{2} + \epsilon_2 \atop 2 \epsilon_5
  }\right](- 2 \xi^2_a ; 2 \tau_2)\right) \ \prod_{q=1}^{N_{D5_1}}
\vartheta^2 \left[{\frac{1}{2} + \epsilon^2 \atop
    \frac{1}{2} +  \epsilon^5 }\right] (\phi^2_q ; \tau_2) \ , \nonumber \\
&& f_3^{(i_3)} ({\xi_a^3, \xi^3_k,\tau_3, N_{D9}}) = \sum_{\epsilon_3,\epsilon_6}
e ^{- 2 i \pi [\epsilon_3 \epsilon_6 + N_{D9} (\epsilon_6 +
  \epsilon_3 \epsilon_6)] } \times \nonumber \\
&& \left(\vartheta\left[{\frac{i_3}{2} + \epsilon_3 \atop
 2 \epsilon_6 }\right](2 \xi^3_a ;  2\tau_3) +
\vartheta\left[{\frac{i_3}{2} + \epsilon_3 \atop 2 \epsilon_6
  }\right](- 2 \xi^3_a ; 2 \tau_3)\right) \ \prod_{k=1}^{N_{D9}}
\vartheta^2 \left[{\frac{1}{2} + \epsilon^3 \atop
    \frac{1}{2} +  \epsilon^6 }\right] (\xi^3_k ; \tau_3) \ .
\label{ff}
\end{eqnarray}
in terms of which the previous wavefunctions are
given by
\begin{eqnarray}
&& \psi^{(i_1,i_1',i_2,i_3)}_1 \ = \  f_1^{(i_1i'_1)} ({\xi_a^1,\tau_1}) \ f_2^{(i_2)} ({\xi_a^2,
  \phi^2_q, \tau_2, N_{D5_1}}) \ f_3^{(i_3)} ({0, \xi^3_k,\tau_3, N_{D9}}) \ ,
\label{psi1} \\
&& \psi^{(i_1,i_1',i_2,i_3)}_2 \ = \ f_1^{(i_1i'_1)} ({\xi_a^1,\tau_1}) \  f_2^{(i_2)} ({0,
  \phi^2_q, \tau^2, N_{D5_1}})  \ f_3^{(i_3)} ({\xi_a^3, \xi^3_k,\tau_3,
  N_{D9}}) \ , \label{psi2} \\
&& \psi^{(i_1,i_1',i_2,i_3)}_3 \ = \  f_1^{(i_1i'_1)} ({0,\tau_1}) \  f_2^{(i_2)} ({\xi_a^2,
  \phi^2_q, \tau_2, N_{D5_1}})  \ f_3^{(i_3)} ({\xi_a^3, \xi^3_k,\tau_3,
  N_{D9}}) \ , \label{psi3} \\
&& \psi^{(i_1,i_1',i_2,i_3)}_4 \ = \ f_1^{(i_1i'_1)} ({0,\tau_1}) \ f_2^{(i_2)} ({0,
  \phi^2_q, \tau_2, N_{D5_1}}) \ f_3^{(i_3)} ({0, \xi^3_k,\tau_3, N_{D9}}) \
. \label{psi4}
\end{eqnarray}
To prove this we have made use of the identities,
\begin{equation}
\xi^i_{a_M}=\xi^i_{a}\ , \qquad \xi^i_{a_M^*}=\xi^i_{a^*}=-\xi^i_{a}\ , \qquad M=g,f,h\ , \quad i=1,2,3
\end{equation}
In order to write the contributions of the disc and one-loop amplitudes to the
superpotential of the 56 antisymmetrics and 8 symmetrics in this model, we observe that under an orbifold generator
which reverses all the coordinates but the ones of the $k$-th 2-torus,
accordingly to eq.(\ref{orbifold}), the vector $\vec i$ transforms as,
\begin{equation}
i_p \rightarrow |I_{aa^*}^p|-i_p \ \ \ (\textrm{mod }\  |I_{aa^*}^p|)
\ , \qquad i_k \rightarrow i_k \ , \qquad p\neq k
\end{equation}
Then, the  orbifold/orientifold invariant contributions read,
{\footnotesize \begin{align}
F_{A^{1,(i_2,i_3)}}&=\psi^{1,(0,0,i_2,i_3)}\ , &
F_{A^{2,(i_2,i_3)}}&=\psi^{2,(0,0,i_2,i_3)}\ ,\label{hol}\\
F_{A^{3,(i_2,i_3)}}&=\psi^{3,(0,0,i_2,i_3)}\ , &
F_{A^{4,(i_2,i_3)}}&=\psi^{4,(0,0,i_2,i_3)}\ ,\nonumber\\
F_{A^{5,(i_2,i_3)}}&=\psi^{1,(2,2,i_2,i_3)}\ , &
F_{A^{6,(i_2,i_3)}}&=\psi^{2,(2,2,i_2,i_3)}\ ,\nonumber\\
F_{A^{7,(i_2,i_3)}}&=\psi^{3,(2,2,i_2,i_3)}\ ,
&F_{A^{8,(i_2,i_3)}}&=\psi^{4,(2,2,i_2,i_3)}\ ,\nonumber\\
F_{A^{9,(i_2,i_3)}}&=\psi^{1,(1,3,i_2,i_3)}\ ,
& F_{A^{10,(i_2,i_3)}}&=\psi^{2,(1,3,i_2,i_3)}\ ,\nonumber\\
F_{A^{11,(i_2,i_3)}}&=\psi^{1,(3,1,i_2,i_3)}\ ,
& F_{A^{12,(i_2,i_3)}}&=\psi^{2,(3,1,i_2,i_3)}\ ,\nonumber\\
F_{A^{13,(i_2,i_3)}}&=\psi^{3,(1,3,i_2,i_3)}+\psi^{3,(3,1,i_2,i_3)}\
, & F_{A^{14,(i_2,i_3)}}&=\psi^{4,(1,3,i_2,i_3)}+\psi^{4,(3,1,i_2,i_3)}\
, \nonumber\\
F_{S^{1,(i_2,i_3)}}&=\psi^{3,(1,3,i_2,i_3)}-\psi^{3,(3,1,i_2,i_3)}\
, & F_{S^{2,(i_2,i_3)}}&=\psi^{4,(1,3,i_2,i_3)}-\psi^{4,(3,1,i_2,i_3)}\
\nonumber \end{align}}
In particular, substituting (\ref{psi1})-(\ref{psi4}) in these
expressions it is possible to check that
$\mathcal{F}_{S^{Q,(i_2,i_3)}}$
is identically zero, in agreement with the fact that, because of gauge
invariance, the
symmetrics in this model cannot get a linear term in the superpotential.

Therefore, the non-perturbative superpotential reads,
\begin{equation}
W_{\rm n.p} \ = \ \left[\eta(\tau^1) \eta(\tau^2)^{3}
  \eta(\tau^3)^{3} \right]^{-1} \  \sum_{i_2,i_3=0,1}\sum_{M=1}^{14}
e^{2\pi i \hat T_3}\hat{A}^{M,(i_2,i_3)}\ F_{A^{M,(i_2,i_3)}}
\end{equation}


\end{document}